\documentclass[fleqn,usenatbib]{mnras}

\usepackage{newtxtext,newtxmath,pdflscape}

\usepackage[T1]{fontenc}
\usepackage{ae,aecompl}
\usepackage{rotating}
\usepackage{amsmath}	
\usepackage{amssymb}	
\usepackage{MnSymbol}
\usepackage{lscape}
\usepackage{colortbl}    
\usepackage{float}
\usepackage{ulem}
\usepackage{color}
\usepackage[dvipsnames]{xcolor}
\usepackage{rotating}

\definecolor{Azure}{rgb}{0.0, 0.5, 1.0}

\newcommand{\hh} {H$_2$}
\newcommand{\mhh} {$M_{\mathrm {H_2}}$}
\newcommand{\HI}{\textsc{Hi}\,}
\newcommand{\Mo}{{\mathrm M}_\odot}

\newcommand{\hmpc}{h^{-1}\,\mathrm{Mpc}}
\newcommand{\hkpc}{h^{-1}\,\mathrm{kpc}}
\newcommand{\dynp} {D9$^+$}
\newcommand{\ftwo}{f_{{\mathrm{H}_2}}}

\def\simlt{\lower.5ex\hbox{\ltsima}}

\title[Comparing $\text{H}_2$ formation models at high $z$]{A comparison of $\text{H}_2$ formation models at high redshift}

\author[Sch\"abe et al.]{Alexander Sch\"abe$^{1,}$\thanks{E-mail: aschaebe@astro.uni-bonn.de}, Emilio Romano-D\'iaz$^{1}$, Cristiano Porciani$^{1}$,
\newauthor Aaron D. Ludlow$^{2}$ and Matteo Tomassetti$^{3}$
\\
$^{1}${Argelander Institut f\"ur Astronomie , Auf dem H\"ugel 71, D-53121 Bonn, Germany}\\
$^{2}${International Centre for Radio Astronomy Research, University of Western Australia, 35 Stirling Highway, Crawley,}\\
  {Western Australia, 6009, Australia}\\
$^{3}${Marks and Spencer Group plc. Waterside House 35 North Wharf Road London W2 1NW}
}

\date{Accepted 2020 July 27. Received 2020 July 3; in original form 2020 March 7}

\pubyear{2020}

\begin{document}
\label{firstpage}
\pagerange{\pageref{firstpage}--\pageref{lastpage}}
\maketitle
\begin{abstract}
Modelling the molecular gas that is routinely detected through CO observations of high-redshift galaxies constitutes a major challenge for ab initio simulations of galaxy formation.
We carry out a suite of cosmological hydrodynamic simulations to compare three approximate methods that have been used in the literature to track the formation and evolution of the simplest and most abundant molecule, \hh. Namely, we consider: i) a semi-empirical procedure that associates \hh~to dark-matter haloes based on a series of scaling relations inferred from observations, ii) a model that assumes chemical equilibrium between the \hh~formation and destruction rates, and iii) a model that fully solves the out-of-equilibrium rate equations and accounts for the unresolved structure of molecular clouds. 
We study the impact of finite spatial resolution and show that robust \hh~masses at redshift $z\approx 4$ can only be obtained for galaxies that are sufficiently metal enriched in which \hh~formation is fast. 
This corresponds to \hh~reservoirs with masses $M_{\mathrm{H_2}}\gtrsim 6\times 10^9$ M$_\odot$.
In this range, equilibrium and non-equilibrium models predict similar molecular masses (but different galaxy morphologies) while the semi-empirical method produces less \hh.
The star formation rates as well as the stellar and \hh~masses of the simulated galaxies are in line with those observed in actual galaxies at similar redshifts that are not massive starbursts.
The \hh~mass functions extracted from the simulations at $z\approx 4$ agree well with recent observations that only sample the high-mass end.
However, our results indicate that most molecular material at high $z$ lies yet undetected in reservoirs with $10^9<M_{\mathrm H_2}<10^{10}$ M$_\odot$. 
\end{abstract}




\section{Introduction}

Observations of molecular gas at high redshift \citep[see e.g.][for a review]{Carilli-Walter-2013} are shaping our knowledge of the early phases of galaxy formation. To fully appreciate their implications, it is vital to develop a theoretical framework within which the experimental findings can be interpreted. However, ab initio simulations of galaxy formation generally do not resolve the spatial scales and densities (nor capture the physics) that characterize molecular clouds in the interstellar medium (ISM), and therefore fall short of modelling the molecular content of galaxies. 
The need for more sophisticated models is therefore becoming increasingly important, particularly with the advent of the Atacama Large Millimeter/submillimeter Array (ALMA) which has enabled detections of molecular-gas reservoirs at redshifts as high as $z \sim 4$-7 \citep[e.g.][]{riechers+14,capak+15,maiolino+14,decarli+16,bothwell+17, santini+19}.

Opening a window on the molecular Universe also motivates new theoretical efforts to gain insight into how galaxies grow their stellar component.
This requires developing a coherent picture that links molecular gas in the turbulent interstellar medium (ISM) to the various feedback processes that regulate the supply of gas available to form stars.
Stellar nurseries in the Milky Way appear to be associated with dusty and dense molecular clouds. Spatially resolved observations of nearby galaxies show that the surface density of star formation (SF) better correlates with the surface density of molecular gas than with the total gas density \citep[e.g.][]{wong&blitz,kennicutt+07,leroy+08,bigiel+08}. A possible interpretation of these findings is that the presence of molecular material is necessary to trigger SF \citep{Krumholz-McKee-05, Elmegreen-07, Krumholz-McKee-Tumlinson-09}, although other viewpoints are also plausible. One possibility, advocated by \citet{Krumholz+11} and \citet{glover&clark12}, is that \hh~and SF are spatially correlated due to the ability of the gas to self shield from interstellar ultraviolet (UV) radiation. That SF primarily takes place in molecular clouds would, in that case, be coincidental rather than a consequence of some fundamental underlying relation between \hh~and SF.

In numerical simulations, the two scenarios generate different galaxies:
\hh-regulated SF is delayed in the low-metallicity progenitors of a galaxy where dust  and central gas densities are too low to activate an efficient conversion of \HI into  \hh~\citep{kuhlen+12, jaacks+13, kuhlen+13, thompson+14, tomassetti+15}. The resulting  galaxies are thus characterized by lower stellar masses, younger stellar populations,  and a smaller number of bright satellites \citep{tomassetti+15}. In addition, the fact  that the energy due to stellar feedback is injected at different locations gives rise to different galaxy morphologies \citep{tomassetti+15, pallottini+17}.

In the ISM, \hh~primarily forms due to the catalytic action of dust grains and is destroyed by resonant absorption of photons in the Lyman and Werner (LW) bands. 
This is why \hh~is abundant in the densest and coldest regions of the ISM where far-UV radiation is heavily attenuated \citep{Draine78, Hollenbach-McKee79,vanDishoeck+Black-86,Black+vanDishoeck-87, Draine+Bertoldi-96, Sternberg-88}.
The main difficulty in tracking molecular gas within galaxy formation models is the huge dynamic range between the scales that tidally torque  galaxies and those that regulate the turbulent ISM and on which SF and stellar feedback take place.

One way to overcome this limitation is to use empirical laws inferred from observations in order to predict the abundance of molecular gas within galaxies.
For example, the ratio between the surface densities of molecular and atomic hydrogen is found to scale quasi-linearly with the interstellar gas pressure in the mid-plane of disc galaxies
\citep{wong&blitz,Blitz+Rosolowsky04,blitz+06, leroy+08}, a fact that has been exploited to develop semi-analytic models (SAMs) of galaxy formation
\citep{Dutton+VDB09, obreschkow+09,obreschkow&rawlings+09_2,fu+10, lagos+11, fu+12, popping+14, popping+15, somerville+15,
lacey+16,stevens+16,lagos+18} and numerical simulations \citep{murante+10, murante+15, diemer+18} that take \hh~into account.

A second possibility for tracking \hh~is to consider a number of simplifying assumptions under which the coupled problems of radiative transfer and \hh~formation in the ISM can be solved analytically \citep[see, e.g.][]{krumholz+08,krumholz+09,mckee&krumholz10}.
In this case, the processes regulating the molecular gas fraction are assumed to be in local equilibrium and the resulting \hh~abundance depends only on the column density and metallicity of the gas.
Model predictions for the \HI-\hh~transition profiles appear to be consistent with observations in external galaxies with different metallicities \citep{Fumagalli+10, Bolatto+11, Wong+13}.
Equilibrium models have been widely used to predict the molecular content of galaxies in the semi-analytic framework \citep{fu+10,lagos+11, fu+12, krumholz&dekel+12, somerville+15} and in numerical simulations of small-to-intermediate cosmological volumes \citep{kuhlen+12,jaacks+13,kuhlen+13, hopkins+14,thompson+14, lagos+15, dave+16}. 
A more complex equilibrium model in which the dust abundance and grain-size distribution evolve with time has been recently employed in simulations of an isolated disc galaxy \citep{chen+18}.

The third option on the market is to model the out-of-equilibrium evolution of the \hh~abundance. The main motivation for doing this is that the formation of \hh~on dust-grains can be a slow process and the chemical rate equations reach equilibrium only if the ISM presents favourable conditions (e.g. high dust content and long dynamical timescales). As a result, the equilibrium models described above may over-predict the abundance of \hh~in certain scenarios. To overcome this problem, one can directly integrate the system of chemical rate equations without resorting to approximate equilibrium solutions. This approach, however, requires accounting for the complex interplay between velocity and density in a turbulent medium that ultimately determines the column density of the gas and dust. While such a line of attack characterizes state-of-the-art simulations of small ISM patches \citep[see, e.g.,][and references therein]{Seifried+17}, it cannot yet be fully implemented in cosmological simulations of galaxy formation as they do not yet resolve the relevant length, time and density scales.
The simplest approach is to solve the chemical rate equations after coarse-graining them at the level of the single resolution elements and introduce a clumping factor in the \hh~formation rate to account for unresolved density fluctuations. The best possible spatial resolution is then achieved by focusing on idealized \citep{pelupessy+06, robertson+08, pelupessy+09, hu+16, richings&schaye16, lupi+18} or cosmological simulations of individual galaxies \citep{gnedin+09, feldmann+11, christensen+12, katz+17, pallottini+17, nickerson+18, lupi+19, pallottini+19}. 
Alternatively, physics at the unresolved scales can be dealt with by introducing a sub grid model that takes into account the probability distribution of local densities and the temperature-density relation obtained in high-resolution simulations of the turbulent ISM. In this case, a 1D slab approximation is used to associate an optical depth to each microscopic density \citep[see][for details]{tomassetti+15}.
Such a model yields \hh~fractions (as a function of the total hydrogen column density) that are in excellent agreement with observations of the Milky Way and the Magellanic Clouds.

Given this variety of techniques, it is worthwhile identifying which regimes, if any, the outputs of simulations based on empirical, equilibrium and non-equilibrium models for the \hh~abundance give consistent results. For instance, \citet{feldmann+11} show that a tight relation between the \hh~fraction and the ISM pressure  emerges naturally in simulations where the chemical rate equations are integrated without assuming local equilibrium. The slope and the amplitude of the relation depend sensitively on the local ISM properties, in particular on the dust-to-gas ratio. 
When the conditions of the ISM are tuned to those of the solar neighbourhood, the resulting correlation closely matches that observed in local galaxies. Moreover, \citet{krumholz&gnedin11} find that the equilibrium model presented in \citet{krumholz+09} agrees well with time-dependent calculations for a wide range of UV intensities if the \hh~abundance is coarse-grained on scales of $\approx 100$ pc and the ISM metallicity is above $0.01\, Z_\odot$.
For lower values of $Z$, however, the agreement rapidly deteriorates. On the other hand, \citet{MacLow-Glover-12} find that equilibrium models do not, in general, reproduce the results of simulations of the turbulent, magnetized ISM when coarse-grained on scales of $\sim 1-10$ pc.
These results suggest that the level of agreement or disagreement between the different approaches depends on the length-scales over which the comparisons are made.

In this paper, we use a suite of cosmological, hydrodynamical simulations to investigate similarities and differences between the three model prescriptions for \hh~formation.
We focus on the molecular content of high-redshift galaxies, similar to those that can be detected with ALMA. In previous work, equilibrium \hh~models have been employed in simulations with widely different spatial resolutions, ranging from the parsec to kiloparsec scales. In order to provide a benchmark for future studies, we therefore investigate how predictions for the molecular content of galaxies are influenced by the spatial resolution of the simulations, focusing on both equilibrium and non-equilibrium models.
To check the reliability of the \hh~models, we also compare the global properties of our simulated galaxies against observations of high-redshift systems, emphasizing differences between the various \hh-formation schemes.

The paper is organized as follows. 
In section~\ref{sec:form_models}, we introduce the semi-empirical, equilibrium and non-equilibrium models used to track the abundance of \hh~in our simulations. Our numerical setup is described in section~\ref{sec:numerical_methods} and finite spatial resolution effects are investigated in section~\ref{sec:results}. 
We compare our numerical results with a series of observational data in section~\ref{sec:comp-observations}. Finally, we conclude providing a summary of our main results in section~\ref{sec:summary}.

\section{Modelling molecular hydrogen}
\label{sec:form_models}

As mentioned above, modelling the \hh~chemistry in simulations of galaxy formation is extremely challenging because it requires simultaneously resolving the very disparate temporal and spatial 
scales relevant for molecular cloud dynamics and galaxy evolution.
An exact treatment of all relevant processes is clearly impossible, yet progress continues to be made both in SAMs and hydrodynamical simulations. 
In this work, we consider three approximate methods that have been previously presented in the literature and are representative of entire classes of models.
This section provides an overview of their most important aspects as well as the relevant details of their implementation.

\subsection{The semi-empirical model (PBP)}
\label{sec:pbp}

As an example of how we can use empirical laws inferred from observations to associate an \hh~mass to a simulated dark-matter halo, we use the method presented by \citet[][hereafter PBP]{popping+15}. 
The model takes, as input, a halo mass and redshift to which a stellar mass and an instantaneous star formation rate (SFR) are assigned using subhalo abundance matching \citep[see][for details]{behroozi+13}. 
Stars and gas are assumed to be distributed according to an exponential profile. 
The scale length of the stellar disc is chosen according to the empirical relation of \citet{vanderwel+14}, while the size of the gaseous
disc is scaled-up by a factor of $2.6$. 
The relative abundance of HI and \hh~is then determined using an empirical scaling with the mid-plane pressure \citep[][]{blitz+06} for an assumed cold gas mass. 
The latter (and the final results for the \hh~mass) are determined iteratively by requiring that the corresponding \hh ~surface density yields a SFR equivalent to that implied by the observed $\Sigma_{\rm H_2}-\Sigma_{\rm SFR}$ relation \citep[as given by][]{bigiel+08}.
By construction, this method yields galaxy gas masses that are consistent with observed SFRs.

\subsection{The equilibrium model (KMT)}
\label{sec:std}

Locally, the \hh~abundance is determined by the competing actions of molecule formation on dust-grain surfaces and dissociation due to the absorption of LW photons.
Provided certain assumptions are made, the local equilibrium abundance of \hh ~can be evaluated analytically \citep{krumholz+08,krumholz+09,mckee&krumholz10}. 
The calculation assumes a spherical molecular cloud shrouded by an isotropic radiation field of LW photons, and an ISM that is in a two-phase equilibrium between the cold and warm neutral mediums. The dust abundance is assumed to scale linearly with the gas metallicity. 
In this scenario, both the UV field intensity and \hh ~fraction depend only on the local column density and metallicity of the gas \citep[see][]{mckee&krumholz10}.

Due to its simplicity, this equilibrium model is commonly employed within SAMs to estimate the relative contributions of atomic and molecular hydrogen to gaseous discs \citep{fu+10,lagos+11, fu+12, krumholz&dekel+12, somerville+15}.
The model can also be implemented in hydrodynamical simulations where, instead, simple estimates of the instantaneous LW radiation field can be used to deduce the local equilibrium \hh ~fraction 
\citep{gnedin&kravtsov11,kuhlen+12,jaacks+13,kuhlen+13, hopkins+14,thompson+14, baczynski+15, lagos+15, tomassetti+15, dave+16}.

\subsection{The dynamical model (DYN)}
\label{sec:dyn}

Our final model tracks directly the time-dependent formation and destruction of \hh ~within the resolution elements of our simulations 
\citep[see][for further details]{tomassetti+15}. 

State-of-the-art numerical simulations of galaxy formation typically reach spatial resolutions of the order of $50$ to $100\,\,{\rm pc}$, comparable to sizes of giant molecular clouds (GMCs). 
Numerical and observational studies of the turbulent ISM, however, indicate that GMCs are rich in substructure on much smaller scales. 
This is often accounted for in cosmological simulations using a gas clumping factor, $C_\rho$ -- an approximation that neglects the complexity of substructure as well as their temperature-density correlations, which may alter \hh ~formation and destruction rates. For that reason, we model the unresolved sub-grid density distribution using a mass-weighted log-normal probability function \citep{kainulainen09,schneider+13}, whose parameters can be determined once a clumping factor has been specified. 
We assume $C_\rho=10$ in all of our simulations since it has been shown to give reasonable results in simulations involving \hh ~\citep{gnedin+09, christensen+12}. 

We adopt a temperature-density relation for unresolved clumps consistent with results from simulations of the turbulent ISM \citep{glover&maclow07}. 
These simulations suggest that the  formation of \hh ~primarily takes place in dense regions where temperatures remain $\lesssim 200 \, \mathrm{K}$. In this implementation, we neglect the collisional destruction of \hh.
Note that we adopt the same value for $C_\rho$ in simulations with linear spatial resolutions that differ up to a factor of four (see section~\ref{sec:suite} for further details). At first sight, this choice might appear to be unphysical as, in the limit of infinite resolution, every small clump should be resolved and $C_\rho \to 1$. Therefore, one expects $C_\rho$ to decrease as the spatial resolution of the simulations increases.
In \citet{dave+16}, for instance, $C_\rho$ is assumed to scale proportionally to the minimum comoving gravitational softening length.
In their implementation of the KMT model, the clumping factor assumes
the values of
30, 15 and 7.5 in simulations with softening lengths of 0.5, 0.25 and $0.125\,h^{-1}$ kpc, respectively.
Note that, continuing this scaling, $C_\rho$ would approach unity if the softening length is further reduced to $\simeq 25$ pc. 
However, in molecular clouds, most of the clumping takes place at `microscopic' scales compared with the size of our simulation cells.

Figure 7 in \citet{2012MNRAS.421.2531M} shows the time evolution of the clumping factor within a simulated molecular cloud in a box of 20 pc (i.e. smaller than our smallest grid cell): $C_\rho$ is always of order 10 for turbulent rms velocities of a few km s$^{-1}$ and even substantially higher in the presence of compressive forcing. Therefore, using a  clumping factor that does not depend on resolution (as we do)
corresponds to assuming that density fluctuations are much more prominent on microscopic length-scales than on scales that are comparable to size of our smallest grid cells.

Recently, \citet{lupi+18} followed the evolution of a
single galaxy at $z=3$ for 400 Myr using a spatially and temporally varying
clumping factor in a simulation with softening lengths of 80, 4 and (up to) 1 pc, for dark matter, stars and gas, respectively. In this case, $C_\rho$ is linked to the subgrid model for the turbulent ISM and assumes median values around 20.

Our assumptions lead to the following system of coupled differential equations describing the formation and destruction of \hh:
\begin{equation}
\frac{\mathrm{d} \langle n_{\text{H}_2} \rangle}{\mathrm{d}t} = \langle \mathcal{R}_{f}(T)\, n_{\rm HI}\, n_{\rm H} \rangle - \langle G\, \kappa\, \Phi_{\mathrm s} \,e^{-\tau} \,n_{\rm H_2} \rangle \,,
\label{eq:mol_hydro_1}
\end{equation}
\begin{equation}
\frac{\mathrm{d} \langle n_{\rm HI} \rangle}{\mathrm{d}t}=-2\, \frac{\mathrm{d} \langle n_{\rm H_2} \rangle}{\mathrm{d}t} \,,
\label{eq:mol_hydro_2}
\end{equation}
with
\begin{equation}
\langle n_{\rm HI} \rangle + 2 \, \langle n_{\rm H_2} \rangle = \langle n_{\rm H} \rangle \,.
\label{eq:mol_hydro_3}
\end{equation}
Here, the brackets $\langle \dots \rangle$ indicate averages taken over the substructure present within a single resolution element of the simulations. They are computed by integrating over the mass-weighted probability density function (PDF) and temperature-density relation specified by the sub-grid model for the turbulent ISM.
The function $\mathcal{R}_{f}$ controls the formation rate of \hh ~on dust grains. The parameter $G$ is the unshielded interstellar UV radiation flux (in Habing units; not to be confused with Newton's constant); $\kappa$ is the photo-dissociation rate of \hh; $\Phi_{\mathrm s}$ is the \hh ~self-shielding function, and $\tau=\sigma_{\rm d} \, N_{\rm H}$ is the dust optical depth in the LW band with column density $N_{\rm H}$ and the photon cross section $\sigma_{\rm d}$, corresponding to a column density $N_{\rm H}=N_{\rm{HI}}+2N_{\rm{H_2}}$. 
Dust abundances are assumed to scale linearly with gas metallicity as $M_\mathrm{dust}\ / M_\mathrm{gas} = 0.008 ~(\mathrm{Z/Z_\odot})$, where $\mathrm{Z_\odot} = 10^{-2}$ \citep{Draine07}. 


\section{Numerical methods}
\label{sec:numerical_methods}

\subsection{Simulation setup}
\label{setup}

We use the adaptive-mesh-refinement (AMR) code \textsc{Ramses} \citep{teyssier02} to run a suite of hydrodynamical simulations.
The code uses a second-order Godunov scheme to solve the hydrodynamic equations, while trajectories of DM and stellar particles are computed using a multigrid Particle-Mesh solver.

We consider a cubic periodic box with a comoving sidelength of 12 $\hmpc$ and assume a cosmological model consistent with the \citet{planck14} results: $\Omega_{\Lambda}=0.692$, $\Omega_{\mathrm{m}}=0.308$, $\Omega_{\mathrm{b}}=0.0481$ and a present-day value of the Hubble parameter of $H_0 = 100\,h$ km s$^{-1}$ Mpc$^{-1}$ with $h=0.678$.

Initial Conditions (ICs) are generated using the \textsc{Music} code \citep{hahn&abel11} at different spatial resolutions but use the same phases and amplitudes for mutually-resolved modes. 
Our high-resolution ICs are imposed on a grid of $512$ cells per dimension, corresponding to a comoving Lagrangian spatial resolution of $\approx 23.4~\hkpc$. In all cases, the ICs are set at redshift $z=99$ and assume a $\Lambda$CDM model with primordial spectral index $n_s = 0.9608$ and a linear rms density fluctuation in 8 $\hmpc$ spheres of $\sigma_8 = 0.826$. 

For each resolution, we carry out two types of simulations:
one following the evolution of collisionless DM alone and another following the co-evolution of DM and baryons. 
The former is used to establish a refinement strategy and to identify the maximum level of refinement achieved during the simulation. 
In principle, grid refinements are based on the standard `quasi-Lagrangian' criterion, i.e. they are triggered if the number of DM particles in a cell exceeds eight or if the baryonic mass is $> 8 \, m_{\mathrm DM}\, \Omega_{\mathrm b} / (\Omega_{\mathrm m} - \Omega_{\mathrm b})$.
However, in order to prevent runaway refinements at early times, we demand that new levels are only triggered at certain times as described, e.g., in \citet{scannapieco+12}. This ensures that the grid resolution in physical units stays approximately constant (although not continuously but in a series of distinct steps).
Moreover, in the hydrodynamic runs, we make sure that the maximum level of refinement for the DM component does not exceed that reached in the DM-only simulations at the same cosmic time. 
On the other hand, the grid for the gas component is allowed to reach one or two additional levels (see section \ref{sec:suite} for details).

We adopt an equation of state with polytropic index $\gamma = 5/3$ for the gas component. 
To avoid spurious fragmentation, thermal pressure is added where needed by increasing the gas temperature so that the Jeans length is resolved with at least four grid cells \citep{truelove+97,Teyssier+10}.

\subsection{Star formation and stellar feedback}
\label{sec:sf}

Simulations of galaxy formation do not resolve the time and length-scales on which SF occurs in the ISM. Therefore, SF needs to be treated in a simplified way on scales comparable with the spatial resolution. 
It is reassuring that the combined action of this rather crude modelling and of stellar feedback in the simulations leads to the emergence of regularities on kpc scales e.g. the Kennicutt-Schmidt relation \citep{schmidt59, kennicutt89, kennicutt98} and global gas depletion times that are in good agreement with observations \citep[see e.g.][and references therein]{agertz-kravtzov16, Orr+17, Semenov+18}.

Following a standard procedure, we impose that SF only takes place within gas cells that i) are part of a convergent flow and ii) have a temperature $T<10^4\, {\mathrm K}$. 
However, we use two different approaches to model SF.
In the simulations based on the KMT model, we impose that SF only takes place where the number density of hydrogen atoms exceeds $n \sim 1\, \mathrm{cm}^{-3}$. 
The selected gas elements with mass density $\rho_{\mathrm{gas}}$ are then converted into star particles according to a stochastic Poisson process with density 
\begin{equation}
\overset{\cdot}{\rho}_{\mathrm{ SF}} = \epsilon\, \frac{\rho_{\mathrm {gas}}}{t_{\mathrm{ ff}}} \;,
\label{eq:sf}
\end{equation}
where $t_{\rm ff} = \sqrt{3\, \pi / (32\, G_{\mathrm N} \, \rho_{\mathrm{ gas}}})$ is the free-fall time of the gas (here $G_{\mathrm N}$ denotes Newton's gravitational constant) and $\epsilon=0.05$ is an efficiency parameter. 

On the other hand, in the runs carried out with the DYN model, we link SF directly to the local \hh ~mass density through the relation
\begin{equation}    
\overset{\cdot}{\rho}_{\rm SF}= \epsilon\, \frac{ \rho_{\mathrm {H_2}}}{t_*} \;,
\label{eq:h2-sf}
\end{equation}
without imposing any criterion on $n$.
Note that, in this case, most SF naturally takes place at high $n$ since \hh ~formation is inefficient at low particle densities.
For example, in our runs, nearly 96 per cent of SF occurs in cells with $n>2.5\,\mathrm{cm}^{-3}$ and 55 per cent takes place where $n>100\,\mathrm{cm}^{-3}$.  
Nevertheless, some \hh-rich cells  inevitably fall short of actual GMC densities ($n\gtrsim 100\,\mathrm{cm}^{-3}$).
For this reason, we define $t_*$ as the minimum of a cell's free-fall time and that of a uniform cloud of density $n=100 ~{\mathrm {cm}}^{-3}$ \citep[see][for details]{gnedin+09}.

All simulations include supernova type II feedback and the associated metal enrichment, as well as cooling from H, He and metals \citep{rasera-teyssier06}. The impact of cosmic reionization is modelled using the spatially uniform UV background derived in \citet{haardt&madau12}.
Following \citet{kuhlen+12,kuhlen+13} and \citet{tomassetti+15}, we set a metallicity floor of $10^{-3} ~\mathrm{Z}_{\odot}$ at $z=9$. This approximately compensates for chemical enrichment from early generations of unresolved SF \citep[e.g.][]{wise+12} and seeds the initial formation of \hh. Self-shielding of dense gas is approximated by exponentially suppressing UV heating in cells where the gas density exceeds $ n_{\mathrm H} \simeq 0.014\ {\mathrm{ cm}}^{-3}$ \citep{tajiri&umemura98}. 

\subsection{Local UV radiation field}
\label{sec:uv}

The KMT and DYN models need an estimate of the intensity of LW radiation in each resolution element of the simulations. We compute this quantity following the approach of \citet{tomassetti+15}.
We model each stellar particle as a population of stars with masses distributed according to a \citet{kroupa01} initial mass function (IMF) and with luminosities consistent with the \textsc{Starburst}99 templates \citep{leitherer+99}. 
We then calculate the total UV luminosity (in the LW band) of the stellar particles as a function of their ages. Finally, we propagate the photons isotropically from the stellar particles assuming that the ISM transitions  abruptly from optically thin to thick at some characteristic length-scale, $r_{\rm LW}$, which corresponds to the size of a few cells.
Note that this includes a geometric dilution following the inverse-square law \citep[more sophisticated but time consuming approaches solve the  radiative transfer problem for UV radiation on the fly assuming a reduced speed of light, e.g.][]{gnedin&kravtsov11, lupi+18}.


\subsection{Haloes and galaxies}
\label{sec:halo_selection}

In order to identify galaxies and their host haloes in our simulations,
we use the Amiga Halo Finder code \citep[AHF;][]{gill+04,knollmann&knebe09}.
We first locate spherical regions with mean density equal to $200\, \rho_{\mathrm c}(z)$, where $\rho_{\mathrm c}(z)\equiv 3\,H^2(z)/(8\,\pi\, G_{\mathrm N})$ is the critical density of the Universe. We then remove unbound particles by iteratively clipping those whose velocities exceed 1.5 times the local escape speed, $\varv_{\mathrm{esc}}=\sqrt{2 |\phi|}$, where $\phi$ is the local gravitational potential.
Note that we consider all matter components to identify the haloes (for instance, the thermal energy of the gas is also taken into account in the unbinding procedure). 
In what follows, we characterise the haloes based on their position (we associate the halo centre with the densest spot), the total mass of the bound material ($M_{\mathrm h}$), and the maximum distance of a bound mass element from the halo centre ($R_{\mathrm h}$). 

It proves useful to track the evolution of particular haloes through simulation snapshots, or to cross-match them between simulations adopting different SF-recipes or \hh ~models. 
This is done using the DM-component only. 
For simulations with the same initial resolution (e.g. the same $l_{\mathrm {initial}}$), this is trivially carried out by matching particle IDs. 
For simulations with different $l_{\mathrm {initial}}$, we associate a lower-resolution counterpart to each highly resolved halo by minimizing an objective function $d$ that depends on the halo positions ($\boldsymbol{x}$), masses and the maximum value of their rotation curves ($\varv_{\mathrm{max}}$): 
\begin{equation} 
\begin{split}
d =  \left[(\boldsymbol{x}_{l_1}-\boldsymbol{x}_\mathrm{l_2})^2/R^2_{\mathrm h}(l_1) \right]^{2/3} 
 &+ \log \left(M_{\mathrm{h}, l_{1}}/M_{\mathrm{h}, l_2}  \right) ^{2/3} \\&+ \log \left(\varv_{\mathrm{max}, l_1}/\varv_{\mathrm{max}, l_2}  \right) ^{2/3}\;,
\end{split}
\end{equation}
where the subscripts $l_1$ and $l_2$ refer to the higher and lower values of $l_{\mathrm {initial}}$, respectively \citep{angulo+15}.

We assume that each halo hosts a central galaxy that occupies the spherical region of radius $R_{\mathrm {gal}} = 0.1\, R_{\mathrm h}$ \citep[e.g.][but see \citet{stevens+14}]{scannapieco+12}. 
The stellar and gas mass of the resulting galaxies are rather insensitive to the precise definition of their outer boundary.
Outliers are driven primarily by rare major mergers. 
In this work, we do not consider satellite galaxies hosted by substructures of the main haloes. 

\subsection{The simulation suite}
\label{sec:suite}

Table~\ref{tab:simu_info} summarizes the main characteristics of our hydrodynamic simulations. We adopt a naming convention for the different runs in which the first letter identifies the \hh ~model that has been used (K for KMT and D for DYN), followed by a number indicating its Lagrangian refinement level, $l_{\mathrm {initial}}$, which varies from 7 for our lowest resolution simulation to 9 for our highest. 
For two runs, we allow gas cells to refine up to two levels higher than the maximum level attained in the corresponding DM-only simulations. 
We use the superscript `+' to distinguish these runs from the others.

On top of the simulations listed in Table~\ref{tab:simu_info}, 
we also build a galaxy catalogue based on the PBP model. 
This lists the \hh~mass, SFR, and the stellar mass that the semi-empirical model associates to the central galaxies of the haloes extracted from the D9$^+$ run.
\begin{table*}
	\centering
	\begin{tabular}{ccccccccc}
		\hline
		Simulation & $\rm{H}_2$ model & SF model  &$l_{\rm{initial}}$ & $l_{\rm{max}}$ & $\Delta x(z=4) \ [\rm{pc}] $ & $z_{\rm{min}}$ & $M_\mathrm{DM}  \ [\rm{M_\odot}]$ & $M_{*}\ [\rm{M_\odot}]$\\
		\hline
		$\,\,\,\rm D9^+$ & DYN & $\rm{H}_2$ & 9 &16 & 55 & 3.6 & $1.4 \times 10^6$ & $2.0 \times 10^5$ \\
		$\rm D9$ & DYN & $\rm{H}_2$ & 9 &15 & 110 & 2.8 & $1.4 \times 10^6$ & $2.0 \times 10^5$\\
		$\,\,\,\rm D8^+$ & DYN & $\rm{H}_2$ & 8 &15 & 110 & 3.8 & $1.1 \times 10^7$ & $6.3 \times 10^5$\\
		$\rm D8$ & DYN & $\rm{H}_2$ & 8 &14 & 220 & 3.9 & $1.1 \times 10^7$ & $6.3 \times 10^5$\\ \hline 
		$\rm K9$ & KMT & gas & 9 & 14 & 110 & 4.3 & $1.4 \times 10^6$ & $2.0 \times 10^5$\\
		$\rm K8$ & KMT & gas & 8 & 13 & 220 & 4.1 & $1.1 \times 10^7$ & $6.3 \times 10^5$\\
		$\rm K7$ & KMT & gas & 7 & 12 & 440 & 4.1 & $8.7 \times 10^7$ & $2.0 \times 10^6$\\ \hline 
	\end{tabular}
	\caption{Main properties of our simulation suite. The first column assigns a name to each run. The names are formed by a letter that denotes the adopted \hh-formation model (second column) and a number that indicates the spatial resolution of the ICs (fifth column). The simulation box of linear size $L_{\mathrm{box}}$ (fourth column) is initially divided in $2^{3l_{\mathrm{initial}}}$ identical cells. During the evolution, we use the AMR technique to increase the spatial resolution in the high-density regions. The maximum level of refinement achieved by the simulations for the gas component is indicated in the sixth column. Note that a new grid refinement is triggered right after $z=4$ and therefore the values of $l_{\text{max}}$ for the D- and K-series (that have different $z_{\mathrm {min}}$) do not seem to match. For this reason, in the seventh column,	we report the spatial resolution (in physical units) achieved by the simulations at $z=4$ (i.e. before the new refinement is triggered). 
	The symbol `+' at the end of the simulation name highlights those runs in which an extra level of refinement is used for the gas component with respect to a normal run. 
	The adopted SF recipe (third column) has a one-to-one association with the \hh~model: the K runs form stars depending on the total gas density while the D ones by the amount of molecular gas (see section \ref{sec:sf} for further details).
	 The minimum redshift reached by the simulations is given in the eighth column. Finally, the minimum masses of the stellar and DM particles are given in the ninth and tenth columns.
	 }
\label{tab:simu_info}
\end{table*}

\section{Numerical resolution effects on the \hh ~mass}
\label{sec:results}

In this section, we investigate how the resulting \hh ~mass of the simulated galaxies is affected by the finite spatial resolution of the runs.

\subsection{Dynamical model}
\label{sec:dynresults}

\begin{landscape}
\begin{figure}
  \centering
  \includegraphics[width=1.3\textwidth,angle=0]{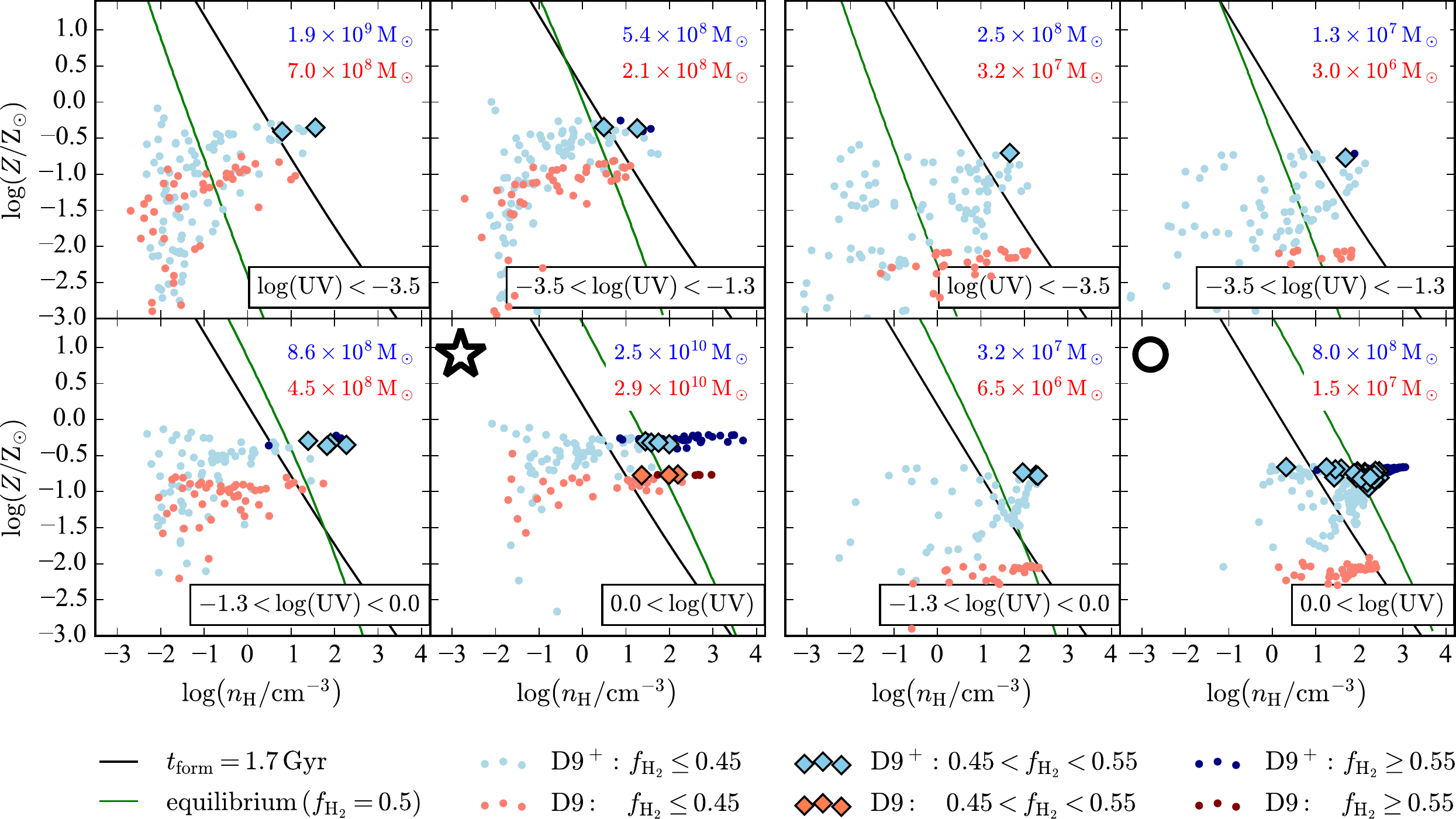}
  \caption{Density, metallicity, and intensity of UV radiation for a subsample of the numerical resolution elements found within two simulated galaxies at $z=3.6$. The four panels on the left-hand side refer to the central galaxy hosted by the most massive halo in our simulations with $M_{\mathrm h}=6.1\times 10^{11}$ M$_\odot$. Those on the right-hand side correspond to the central galaxy hosted by a much smaller halo with $M_{\mathrm h}=4.4\times 10^{10}$ M$_\odot$. The red data points are extracted from the D9 simulation while the blue ones come from the D9$^+$ run.
  Different symbols indicate the \hh~fraction as described in the legend. The numbers in the top-left corners give the total \hh~mass contained in each panel for the two simulations.
  The green and black lines mark the loci where the equilibrium \hh~fraction is 0.5 and where the \hh~formation time equals the age of the Universe, respectively (see the main text for further details). 
  }
	\label{fig:cells_star_n_circle}
\end{figure}
\end{landscape}

In our dynamical model, the net formation rate of \hh~ultimately depends on the gas density, metallicity and the intensity of UV radiation in each resolution element of the simulations. 
In Fig.~\ref{fig:cells_star_n_circle}, we illustrate how the gas cells in the D9$^+$ (blue) and D9 (red) runs populate this space for two particular galaxies at $z=3.6$. 
The first one, denoted with the symbol $\largestar$ and displayed in the left block of panels, is the central galaxy hosted by the most massive halo in our simulations at $z=3.6$, with $M_{\mathrm h}=6.1\times 10^{11}$ M$_\odot$.  
The second one, denoted with the symbol $\bigcirc$ and displayed in the right block of panels, is hosted by a much smaller halo with $M_{\mathrm h}=4.4\times 10^{10}$ M$_\odot$. 
For each galaxy, we consider four bins for the UV intensity and we show the scatterplot of the cells in the $Z$-$n_{\mathrm{gas}}$ plane. 
To improve readability, we only plot one in every 300 cells for the  D9$^{+}$ run and one in every 5 cells for the D9 simulation.
The colour and shape of the symbols indicate whether the \hh~fraction in a cell is $\ftwo\geq$0.55 (the dark circles), $0.45<\ftwo<0.55$ (the diamonds), and $\ftwo\leq0.45$ (the light circles). 
The green lines represent the loci where the \hh~abundance is in equilibrium (i.e. where the formation rate equals the destruction rate  at $\ftwo=0.5$). They are computed assuming the median value of the UV intensity in each panel and a cell size of 59 pc corresponding to the highest refinement level in the D9$^+$ run. Note that they shift towards the right for more intense UV radiation as higher gas densities (at fixed metallicity) are necessary to maintain equilibrium in the presence of an increased destruction rate for the \hh~molecules.
For the lowest UV intensities, equilibrium could in principle be reached at relatively small gas densities. However, the \hh~formation time, $t_{\mathrm{ form}}=[\mathcal{R}_f(T)\, n]^{-1}$, can be extremely long. 
In this case, more time is needed to reach equilibrium.
For this reason, we use black lines to indicate the loci where $t_{\mathrm{ form}}$ equals the age of the Universe at $z=3.6$ (1.7 Gyr). 
In each panel, we expect to find abundant \hh~only on the right-hand side of both the green and black lines.
In other words, the simulations must resolve high-enough densities to produce substantial amounts of \hh. Moreover, the relevant density threshold changes with the local metallicity and UV intensity.

\begin{figure}
  \includegraphics[width=\linewidth]{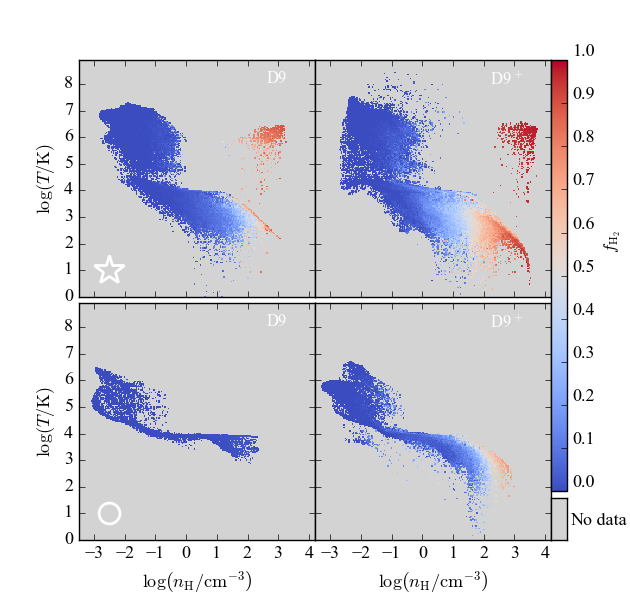}
  \caption{Phase diagram in the density-temperature plane colour coded according to the \hh~density for the gas cells of the galaxies considered
  in Fig.~\ref{fig:cells_star_n_circle}. In the top panels, the small cloud of \hh-rich gas at high densities and temperatures represents elements that have been recently influenced by stellar feedback and that are located around the centre of the galaxy.}
	\label{fig:phase_diagram}
\end{figure}

Let us now focus on the gas cells that form the $\largestar$ galaxy. As expected, in both the D9 and D9$^+$ simulations, the numerical resolution elements that contain large \hh~fractions are generally found on the right-hand side of (or around) the green and black lines.
In the D9$^+$ simulation, cells with large \hh~fractions are found in all bins of UV intensity. 
On the other hand, the fact that the $\largestar$ galaxy is less metal enriched in the D9 run (by $\sim 0.5$ dex) pushes the threshold for copious \hh~formation to higher densities. 
In consequence, \hh~fractions above 0.5 are almost exclusively found in the densest cells (that typically are also associated with larger UV intensities).
To emphasize this difference, in Fig.~\ref{fig:phase_diagram} we show the $n_{\mathrm H}$-$T$ phase diagram of the ISM colour coded by the \hh~fraction.
Since the total \hh~content is dominated by the contribution of these dense gas elements, the two simulations give very similar results for the molecular mass of the $\largestar$ galaxy.

The discrepancy between the D9 and D9$^+$ simulations is more extreme for the $\bigcirc$ galaxy. 
In this case, the metallicity difference between the two runs is larger ($\sim 1$ dex) and, even at the largest resolved densities, the D9 simulation contains very few cells in which $\ftwo \gtrsim 0.5$ while there are many in the D9$^+$ run. 
This happens because $t_{\mathrm{ form}}$ is longer than the age of the Universe for the combinations of densities, metallicities, and UV intensities appearing in the D9 simulation. 
A substantial difference then appears in the predicted total \hh~mass of the $\bigcirc$ galaxy in the D9 and D9$^+$ runs.

The two examples discussed above suggest that 
the earlier onset of SF in higher resolution simulations leads
to faster metal enrichment of the ISM which, in turn, boosts the formation rate of \hh. 
This chain of events is amplified because we link SF to the local \hh~density as explained in section~\ref{sec:sf} as no
SF and metal enrichment can take place before some \hh~is formed in the first place.
Once the metallicity of the ISM is sufficiently large and the timescale for \hh~formation is sufficiently short at the resolved densities, we expect simulations with different resolutions to yield similar \hh~masses.
At the high redshifts we are investigating here, this basically implies that the difference in metallicity between the simulations is not too large.

\begin{figure*}
 \includegraphics[width=\linewidth]{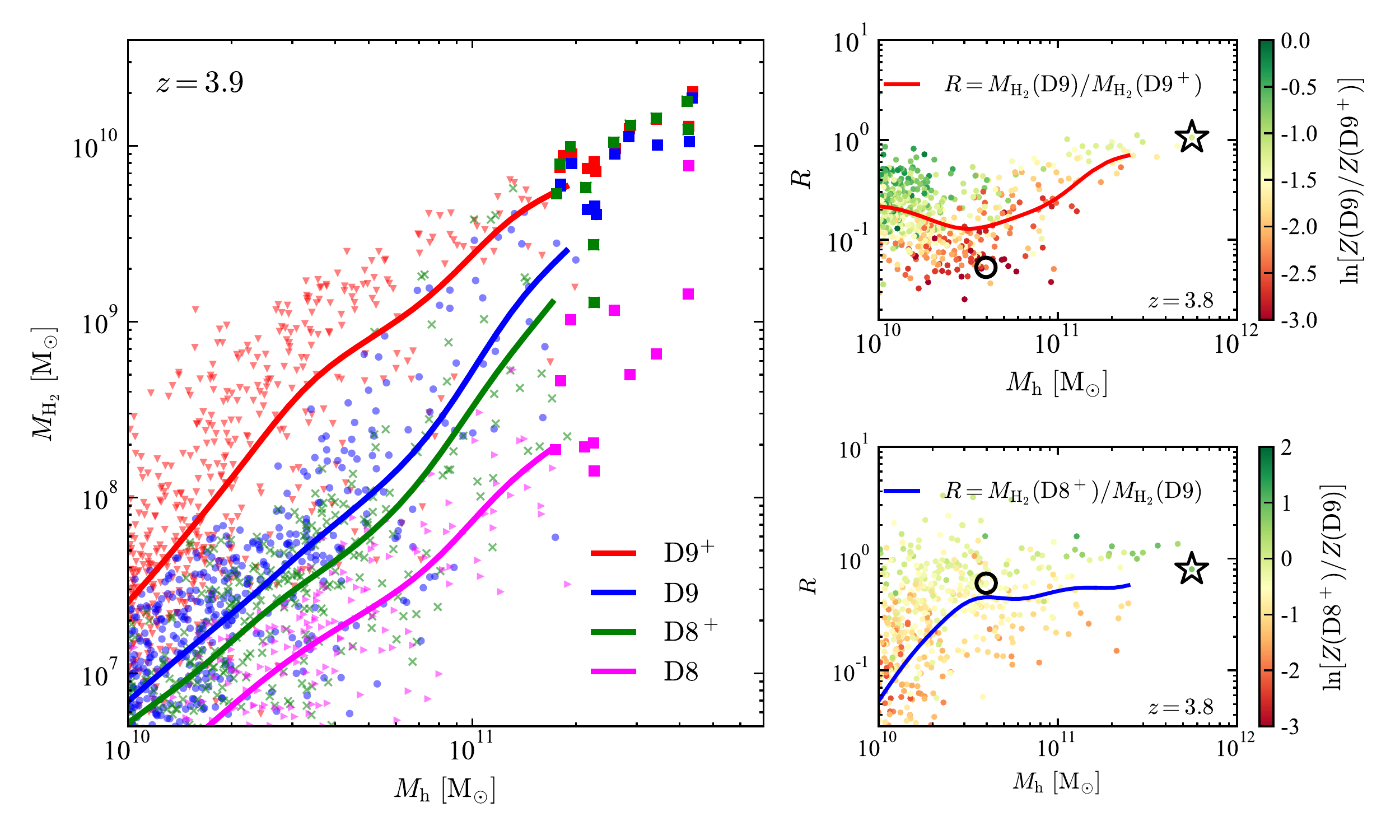}
  \caption{Dependence of the \hh~mass of a galaxy on halo mass and on the maximum refinement level achieved in the DYN simulations. Left: Each symbol shows the \hh~mass of a central galaxy in one of the ``D`` runs as a function of the mass of the corresponding host halo at $z=3.9$.
  The colours distinguish simulations with different maximum spatial resolutions as indicated by the labels. The solid lines represent the running average of the points computed in log-log space using a Gaussian kernel with a standard deviation of 0.1 dex. 
   The square symbols highlight the 11 galaxies with the highest $M_{\mathrm H_2}$ in the D9$^{+}$ simulation and their counterparts in the other runs. These objects will be further discussed in Fig.~\ref{fig:selec_11} and section~\ref{sec:comp-observations}.
  Right: The dots represent the ratio between the \hh~masses of individual objects cross-matched between the D9 and D9$^{+}$ simulations (top) and the D8$^{+}$ and D9 runs (bottom) at $z=3.8$ as a function of the host-halo mass.  
  All points are colour coded based on the (natural) logarithmic difference in metal abundance between the lower- and higher-resolution runs. The outsized star and circle highlight the galaxies discussed in Fig.~\ref{fig:cells_star_n_circle}. The solid lines show running averages taken as in the left panel.
  }
	\label{fig:dyn_models}
\end{figure*}
We now proceed to study the impact of the finite spatial resolution on the global galaxy population. The scatter plot in the left panel of Fig.~\ref{fig:dyn_models} displays the relationship between the \hh ~mass ($M_{\mathrm H_2}$) and the total host-halo mass  ($M_{\mathrm h}$) emerging for individual galaxies in the \dynp (red),  D9 (blue), D8$^{+}$ (green) and D8 (magenta) simulations. We only consider the central galaxies of haloes with $M_{\mathrm h}>10^{10} \, h^{-1} \, \mathrm{M}_\odot$
at the lowest common redshift of the simulations ($z=3.9$).
To highlight the characteristic trends we represent the running averages (computed in log-log space using a Gaussian filter with a standard deviation of 0.1 dex) with solid lines. 

Simulations that achieve the same maximum spatial resolution for the baryonic component (i.e. D8$^+$ and D9) produce very similar results over the entire range of $M_{\mathrm h}$. The extra refinement level in the ICs of the D9 simulation causes only a minor systematic shift of  $M_{\mathrm H_2}$ towards higher values. On the other hand, the galaxies in the D9$^+$ run contain substantially larger \hh~reservoirs at low and intermediate $M_{\mathrm h}$.  It is only for the most massive haloes (for which we do not plot the running averages as they would be biased low) that the D8$^+$, D9 and D9$^+$ simulations yield consistent values of $M_{\mathrm H_2}$.

We further explore these trends in the top-right hand panel of Fig.~\ref{fig:dyn_models} where we plot the ratio of the \hh~masses found in the D9 and D9$^+$ simulations for each central galaxy as a function of $M_{\mathrm h}$ of the host halo (which practically does not change among the various runs).
The solid line once again denotes the Gaussian-weighted running average. In order to connect this statistical study with the detailed discussion, we have presented for the $\largestar$ and $\bigcirc$ galaxies in Fig.~\ref{fig:cells_star_n_circle}; we make sure that the colour of each data point reflects the ratio between the median mass-weighted metallicity of the ISM in the two runs. We also highlight the $\largestar$ and $\bigcirc$ objects themselves with the corresponding symbols. The plot clearly shows that galaxies hosted by haloes with $M_{\mathrm h}<4\times 10^{10} \, h^{-1} \, \mathrm{M}_\odot$ in the D9 simulation tend to contain nearly an order of magnitude less \hh ~than in the D9$^{+}$ run.
However, the scatter is large and strongly correlates with the ratio in the metal content of the galaxies in the two simulations.

As expected from our discussion of Fig.~\ref{fig:cells_star_n_circle}, galaxies that produce many more metals in D9$^+$ (reddish data points) also show a large difference in the \hh~content between the simulations. For these objects, the onset of SF in their progenitors takes place at earlier times in the D9$^+$ run (which is able to resolve higher densities) than in D9 \citep[see also][]{kuhlen+13, tomassetti+15} and this ultimately leads to more metal- and \hh-rich galaxies.

It is interesting to comment also regarding the cloud of greenish data points that appear on the left-hand side of the plot. They correspond to galaxies that have experienced little SF in both simulations and thus show similar levels of metal enrichment. The higher densities resolved in the D9$^+$ run, though, are enough to yield slightly larger \hh~masses.

Finally, in the bottom-right panel of Fig.~\ref{fig:dyn_models}, we compare the \hh~content and the metallicity of the galaxies produced in the D8$^+$ and D9 runs at $z=3.8$. 
These simulations achieve the same maximum level of refinement for the gas although the dark-matter distribution is discretized using particles of different masses. 
Also in this case, we find that the \hh~mass ratio strongly correlates with the relative metallicity. However, the overall trend is different than in the top-right panel. Here, there is a sizeable subpopulation of galaxies for which the two simulations give consistent results at all halo masses. In parallel, there is a second subset whose elements generate substantially less metals and \hh~in the D8$^+$ run. The dichotomy is produced by the absence or presence of a time delay between the epochs in which the progenitors of the galaxies start forming \hh~and stars in the two simulations. 

\begin{figure}
  \includegraphics[width=\linewidth]{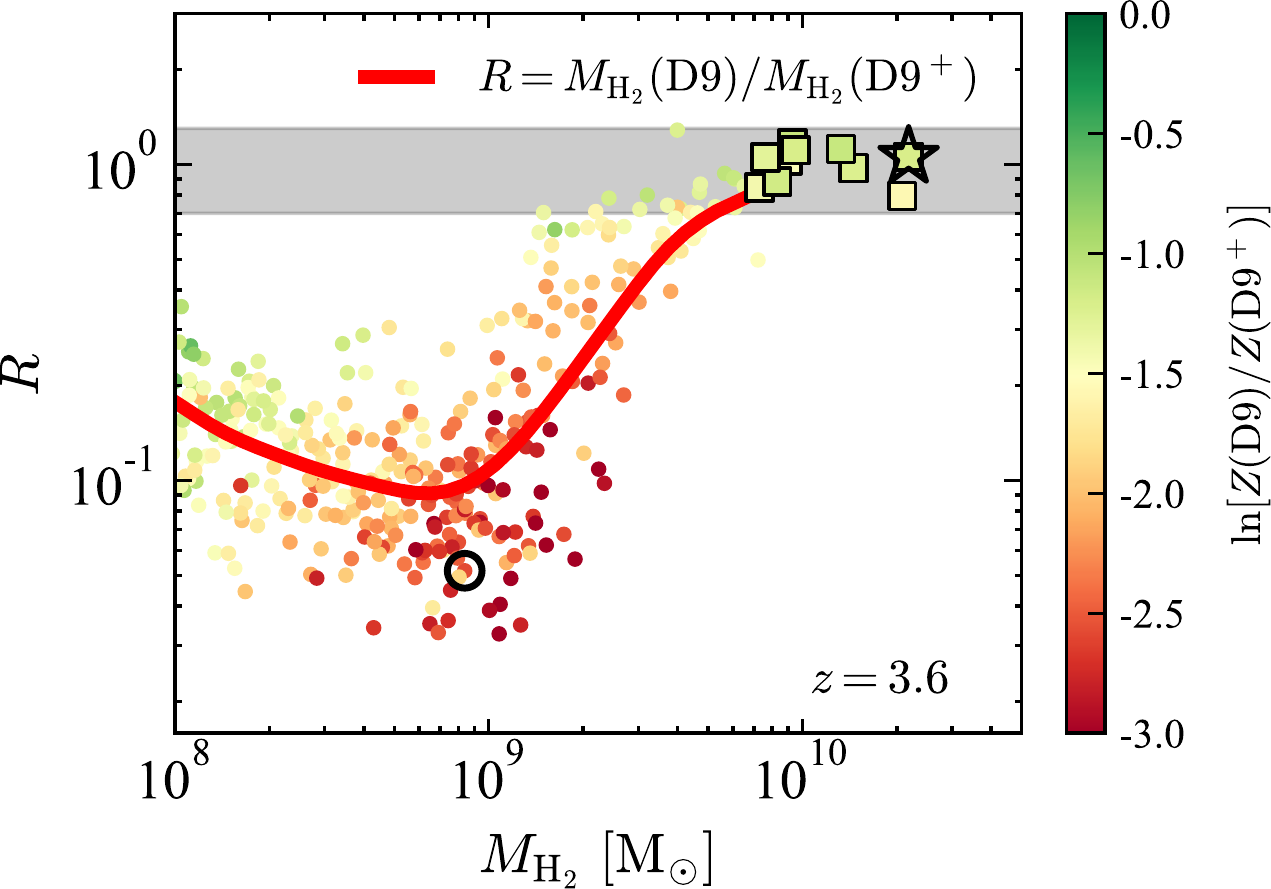}
  \caption{As in the top-right panel of Fig.~\ref{fig:dyn_models} but at $z=3.6$ and using the galaxy \hh~masses instead of the host-halo masses. The grey band highlights the region $0.7<R<1.3$. 
  The square symbols mark the galaxies with the highest \hh~content that lie within the grey band. We will use these 11 objects in section~\ref{sec:comp-observations}.
  }
    \label{fig:selec_11}
\end{figure}

To compare our simulations with observational data and make predictions for forthcoming surveys, we isolate a set of galaxies whose \hh~content does not appear to be strongly affected by spatial resolution effects. 
In practice, we fix a threshold in $M_{\mathrm H_2}$ such that, for all galaxies above it, the \hh~masses found in the D9 and D9$^+$ runs differ by less than 30 per cent at $z=3.6$.
As shown in Fig.~\ref{fig:selec_11}, following this procedure, 
we end up selecting 11 central galaxies with $M_{\mathrm H_2}\geq 10^{9.8}$ M$_\odot$. We will use this subsample in section~\ref{sec:comp-observations}. 

\begin{figure}
  \includegraphics[width=\linewidth]{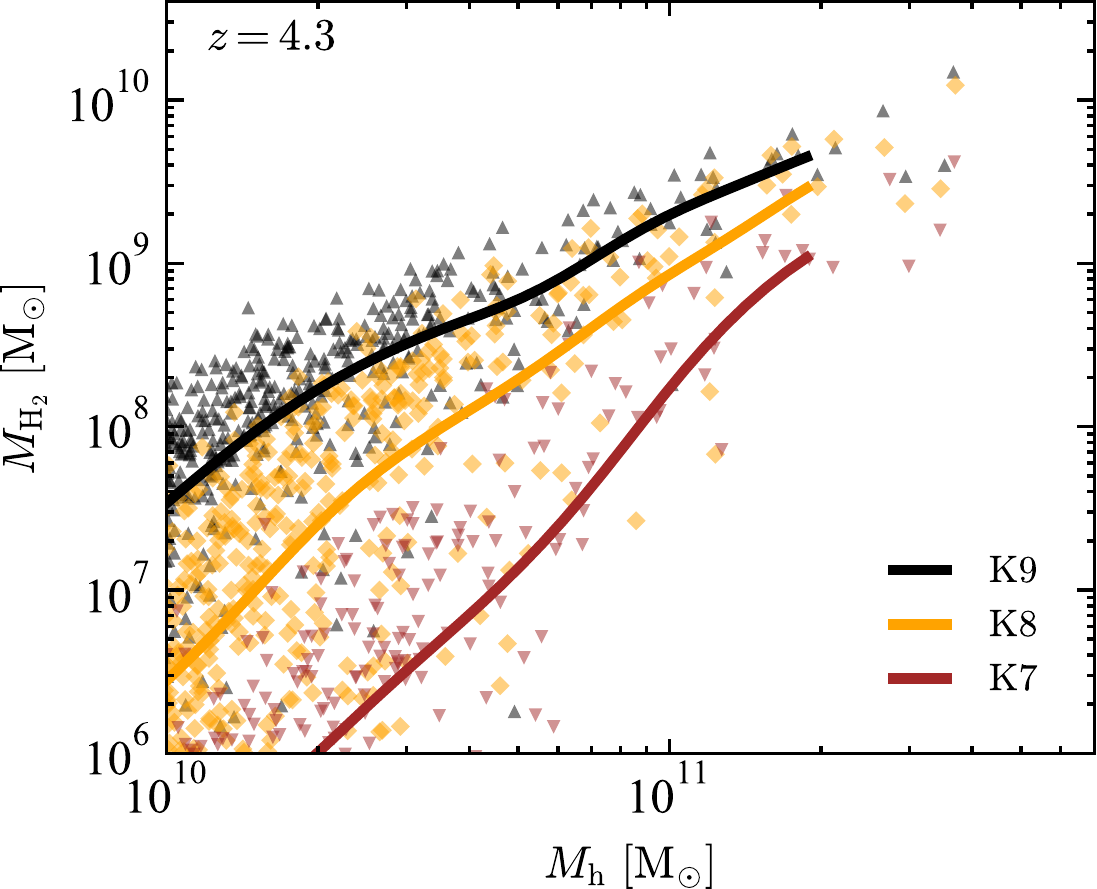}
  \caption{Dependence of the \hh~mass of a galaxy on halo mass, for different maximum refinement levels achieved in simulations based on the equilibrium model. The plot is 
  as in the left panel of Fig.~\ref{fig:dyn_models} but using the K7, K8 and K9 simulations at their lowest common redshift, $z=4.3$.}
	\label{fig:k_models}
\end{figure}

\subsection{Equilibrium model}

The $M_{\mathrm H_2} - M_{\mathrm h}$ relation emerging in the K7, K8 and K9 runs is shown in Fig.~\ref{fig:k_models} at the lowest common redshift of the simulations, $z=4.3$. 
The central galaxies hosted by haloes with $M_{\mathrm h}\simeq 10^{10}$ M$_\odot$ in the K9 simulation contain nearly a factor of 10 (100) more \hh~than in the K8 (K7) run.
This systematic discrepancy decreases with increasing the halo mass. For instance, the difference between K9 and K8 reduces to a factor of $\sim 2-3$ for $M_{\mathrm h}\sim 3 \times 10^{10}$ M$_\odot$ and becomes very small for $M_{\mathrm h}> 2 \times 10^{11}$ M$_\odot$. This trend and its underlying explanation are very similar to those discussed in  Fig.~\ref{fig:dyn_models}.

\begin{figure}
  \includegraphics[width=\linewidth]{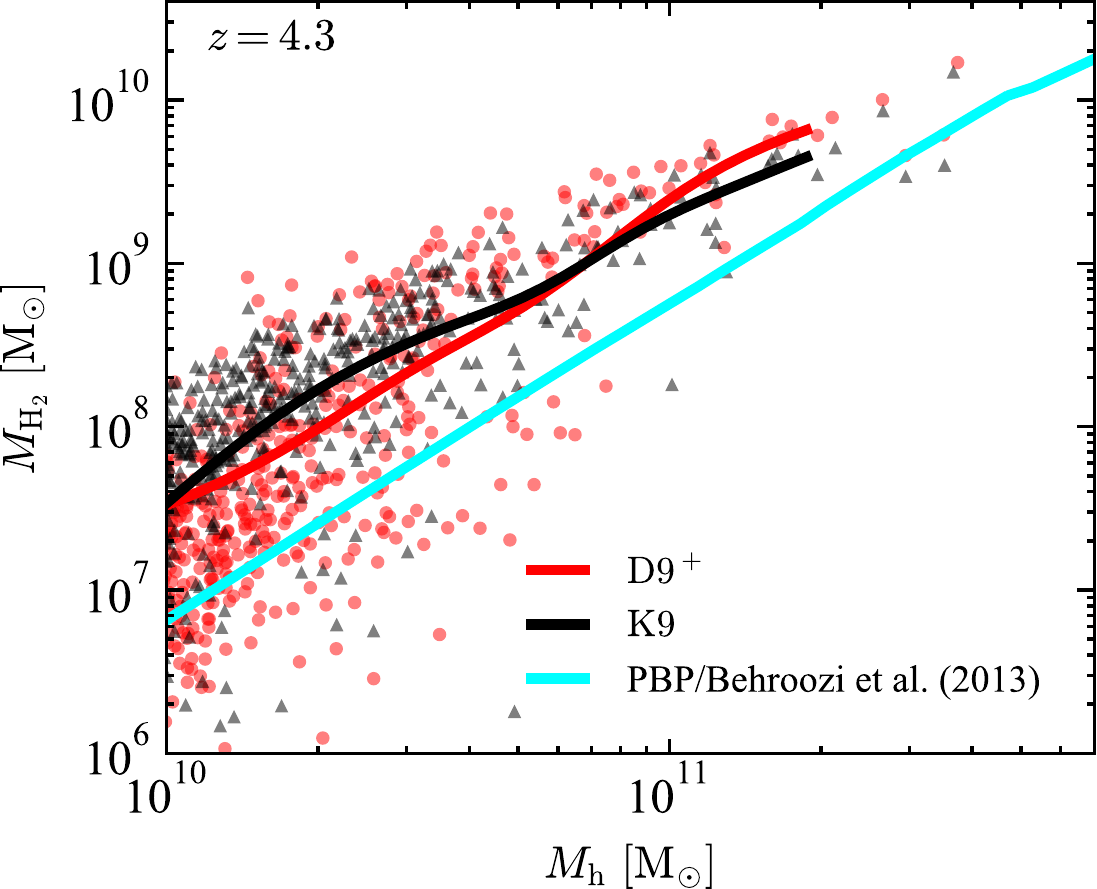}
  \caption{$M_{\mathrm H_2}$-$M_{\mathrm h}$ relation at $z=4.3$ in simulations based on different \hh-formation models. For each algorithm, we use the highest-resolution run available. Note that the \hh~mass in the PBP model only depends on the host-halo mass of the galaxies and so there is no scatter. }
	\label{fig:all_models}
\end{figure}

\begin{figure}
\centering
  \includegraphics[width=0.65\linewidth]{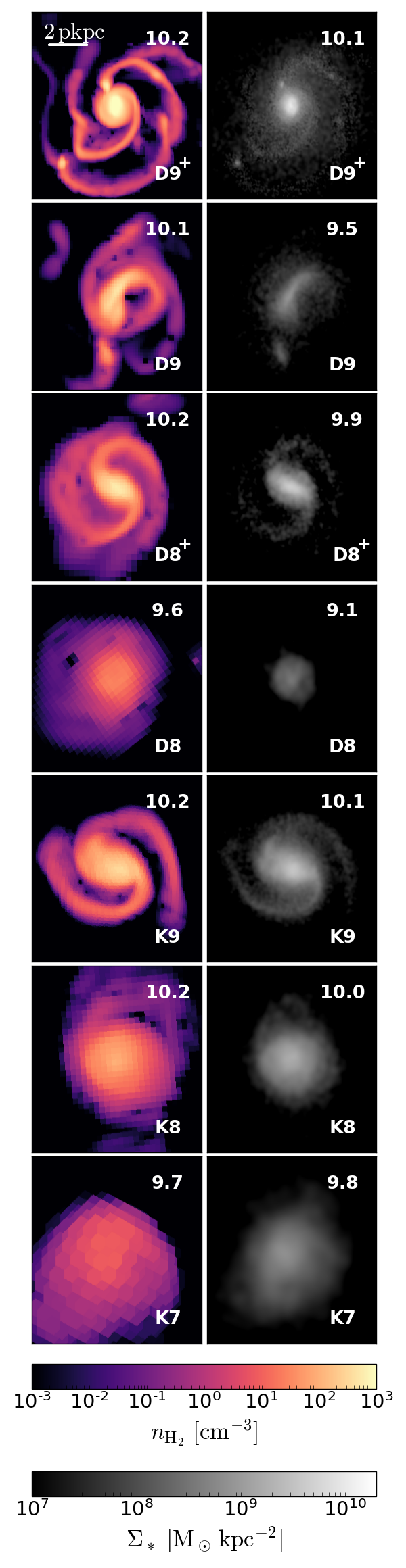}
  \caption{\hh~(left) and stellar (right) face-on maps for the $\largestar$ galaxy in the D9$^+$, D9, D8$^+$, D8, K9, K8 and K7 simulations at $z=4.3$. The images show the maximum \hh~density achieved along the line of sight and the projected stellar density. The white bar in the top-left panel corresponds to a physical length (i.e. not comoving) of 2 kpc. The numbers in the top-right corners give $\log(M_{\rm H_2}/{\mathrm M}_\odot)$ and $\log(M_{\rm *}/{\mathrm M}_\odot)$. }
  \label{fig:h2_maps}
\end{figure} 

\subsection{Model comparison and \hh~maps}
\label{sec:o-2-o}

It is interesting to compare the different models at the highest resolution available. In Fig.~\ref{fig:all_models}, we present the $M_{\mathrm H_2} - M_{\mathrm h}$ relation emerging at $z=4.3$ in the D9$^+$, K9 and PBP simulations. On a statistical basis, the results from the D9$^+$ and K9 simulations agree very well. while the PBP model predicts substantially lower \hh~masses. 
If the comparison is performed object by object, the D9$^+$ run always gives slightly larger molecular masses for $M_{\mathrm h}>10^{11}$ M$_\odot$ and tends to yield smaller $M_{\mathrm H_2}$ for $M_{\mathrm h}<5\times 10^{10}$ M$_\odot$. 

As an illustrative example, in the top panels of Fig.~\ref{fig:h2_maps} we show \hh ~maps of the $\largestar$ galaxy obtained at $z=4.3$ with different models and by varying the numerical resolution.
Each galaxy has been independently rotated using the direction of its stellar angular momentum to obtain a face-on view. Shown is the maximum value of the \hh~density along each line of sight. The name of the simulations and the base-10 logarithm of the total \hh~mass are indicated in each thumbnail. The white bar in the top-left corner corresponds to two physical kpc. For completeness, in the bottom panels of Fig.~\ref{fig:h2_maps}, we also show the projected stellar density
and the base-10 logarithm of the stellar mass.

A few things are worth noticing.
First, the total \hh~and stellar masses come out to be in the same ball park for all runs. In particular, the same value of $M_{\mathrm H_2}$ is consistently found in all simulations that have a spatial resolution $\Delta x(z=4.3)\sim 100$ pc or better (i.e. D8$^+$, D9, D9$^+$, K9).
Secondly, the detailed morphological structure of the galaxy depends substantially on the maximum spatial resolution and on the adopted SF law that changes the way stellar feedback influences the gas.
In the simulations with the lowest resolution (D8, K7, K8), the $\largestar$ galaxy takes the form of a featureless disc.
On the other hand, a strong bar/bulge plus symmetric spiral arms in the disc are noticeable in the D8$^+$ and K9 runs. Signs of interactions (with a smaller companion) appear in the D9 simulation. Finally, a grand design spiral with a small bulge is produced in the D9$^+$ run.
Note that \hh~traces the densest regions of the galaxy in all simulations. 

Similar conclusions can be drawn by inspecting any of the 11 selected galaxies with stable predictions for the \hh~mass in the D runs.
Quite interestingly, all these objects present a disc-like morphology and show prominent spiral arms in the higher-resolution runs.

\begin{figure*}
	\centering
  \includegraphics[width=\linewidth]{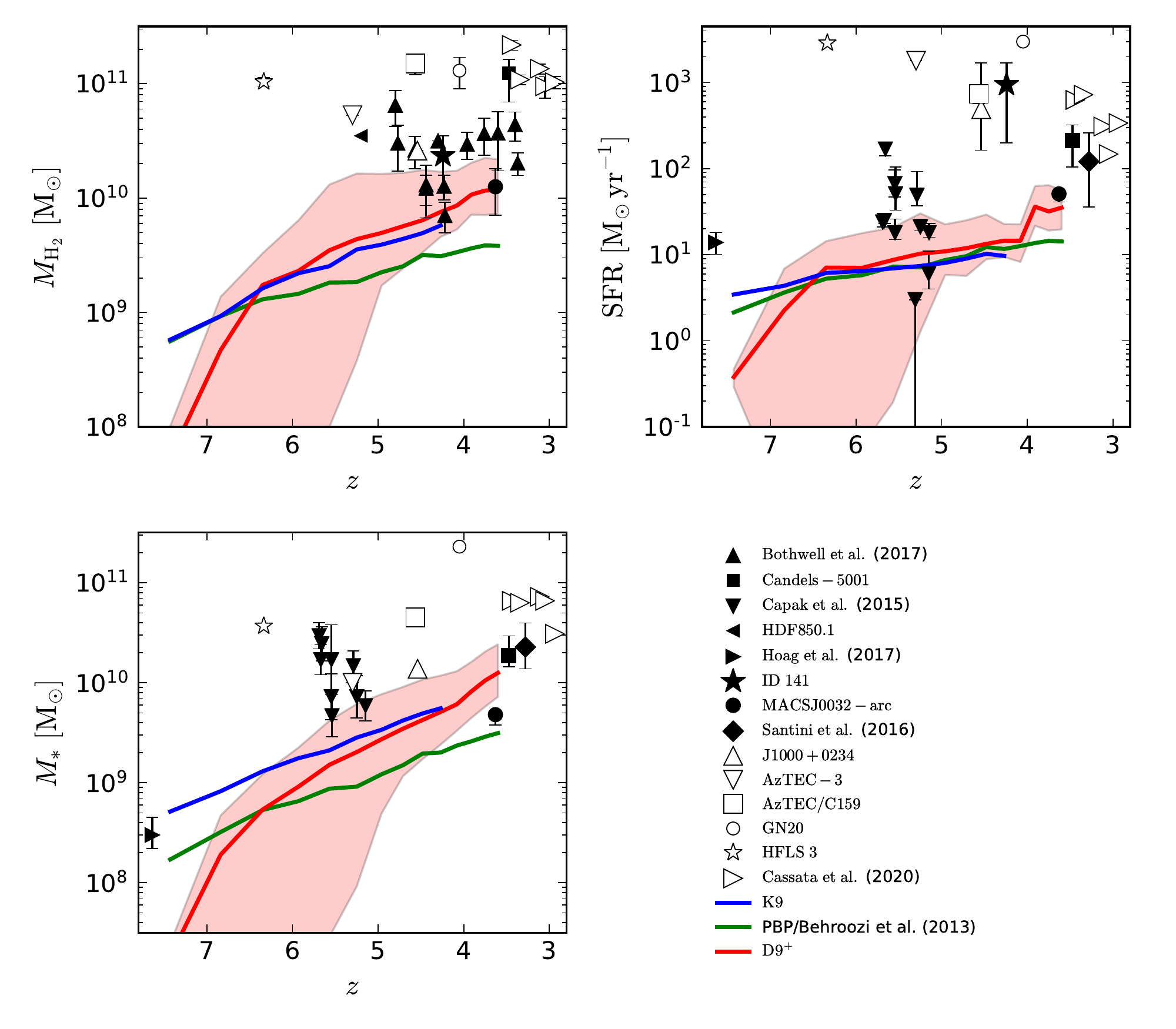}
  \caption{Properties of simulated and real galaxies as a function of redshift.
  The solid lines represent the mean trend of the 11 galaxies with robust predictions for $M_{\mathrm H_2}$ we have selected from our simulations in section~\ref{sec:dynresults}. Their full range of variability in the D9$^+$ run is indicated with the shaded region.
  Symbols with errorbars refer to observed high-$z$ galaxies for which measurements became recently available (see section~\ref{sec:galaxy_properties} for details). Note that the observational sample mainly includes extreme objects that are not representative of the overall galaxy population.}
	\label{fig:results_panel}
\end{figure*}

\section{Comparison with observations}
\label{sec:comp-observations}

In this section, we compare the properties of the 11 galaxies we have selected from our simulations to recent observational data.
First, we look at individual galaxy properties and then we examine the \hh~mass function (MF). Finally, we investigate the time evolution of the cosmic \hh~density.

\subsection{Galaxy properties}
\label{sec:galaxy_properties}
In Fig.~\ref{fig:results_panel}, we contrast the main properties of our simulated galaxies against a compilation of observational data that includes
1) a sample of 13 dusty star-forming galaxies from \citet{bothwell+17} for which [CI] and CO observations are available,
2) Candels - 5001, a single extended object detected in CO \citep{ginolfi+17},
3) 12 dust-poor galaxies detected in [CII] \citep{capak+15}, 
4) ID 141, a sub-millimeter galaxy at $z=4.24$ detected in CO, [CI] and [CII]  \citep{cox+11};
5) HDF 850.1, a galaxy detected at $z\approx 5.2$ \citep{walter+12}, 
6) a spectroscopically confirmed ultra-faint galaxy at $z\sim 7.64$ which presents strong Lyman-$\alpha$ emission \citep{hoag+17},
7) MACSJ0032-arc, a lensed galaxy at $z \sim 3.6$ with detected CO emission \citep{dessauges-zavadsky+17},
8) a lensed sub-millimeter galaxy at $z\sim 3.28$ detected with ALMA \citep{santini+16},
9) J1000+0234, a millimeter galaxy detected in the COSMOS field \citep{schinnerer+08} with SFR and masses obtained by \citet{gomez-guijarro+18},
10) GN20, a sub-millimeter galaxy, member of a rich proto-cluster at $z=4.05$ in the GOODS-North field \citep{carilli+10},
11) AzTEC3, a sub-millimeter galaxy at $z=5.298$ within a massive protocluster in the COSMOS field \citep{riechers+10},
12) HFLS3, a massive starburst galaxy at $z=6.34$ \citep{riechers+13},
13) AzTEC/C159, a star-forming disc galaxy at $z=4.567$ \citep{jimenez-andrade+17},
14) Five star-forming galaxies detected with ALMA using multiple CO transitions and the continuum \citep{Cassata+20}.
It must be acknowledged that this compilation is not representative of the overall galaxy population. 
Only the most luminous objects with extraordinarily high SF rates can be detected at high redshift with current telescopes. This bias becomes even more extreme for the galaxies for which we can estimate the molecular mass. Many of these measurements rely on the amplification of the sources due to gravitational lensing \citep[e.g.,][]{cox+11,walter+12,santini+16,dessauges-zavadsky+17}.

In the different panels of Fig.~\ref{fig:results_panel}, the $\rm H_2$ masses (upper left), stellar masses (lower left), and SFRs (upper right) of the 11 simulated galaxies that have been selected in section~\ref{sec:dynresults} (solid lines and shaded area) are compared with the observational data (the symbols with errorbars) when available.
The solid red lines show the mean values (the actual one and not the average of the log values) for the 11 galaxies extracted from the \dynp run while the shaded regions extend from the minimum to the maximum value.
Similarly, the blue lines indicate the averages for the 11 galaxies extracted from the K9 simulation. 
To improve readability, we do not show their scatter, which is comparable to the shaded region.
Finally, the green lines represent the mean predictions of the PBP model applied to the parent haloes of the \dynp~galaxies. Once again, the corresponding scatter (not shown) is comparable to the shaded region in the plot. 

Fig.~\ref{fig:results_panel} shows that all models predict fast molecular enrichment of the galaxies in the redshift range $3<z<8$. Basically $M_{\mathrm H_2}$ increases by a factor of ten in just a Gyr.
This is associated with a mild increase in the SF rate and a rapid growth of the stellar mass.
On average, the differences between the models are rather small compared with the scatter among the individual galaxies. The most noticeable differences are i) the dynamical model predicts a delayed assembly of the molecular and stellar masses that approximately match the other models only for $z<7$; and ii) the PBP model provides lower estimates  for the molecular (by a factor of 2-3) and stellar masses (by a factor of 2-4) at $z<6$. 

\begin{figure}
	\centering
  \includegraphics[width=\linewidth]{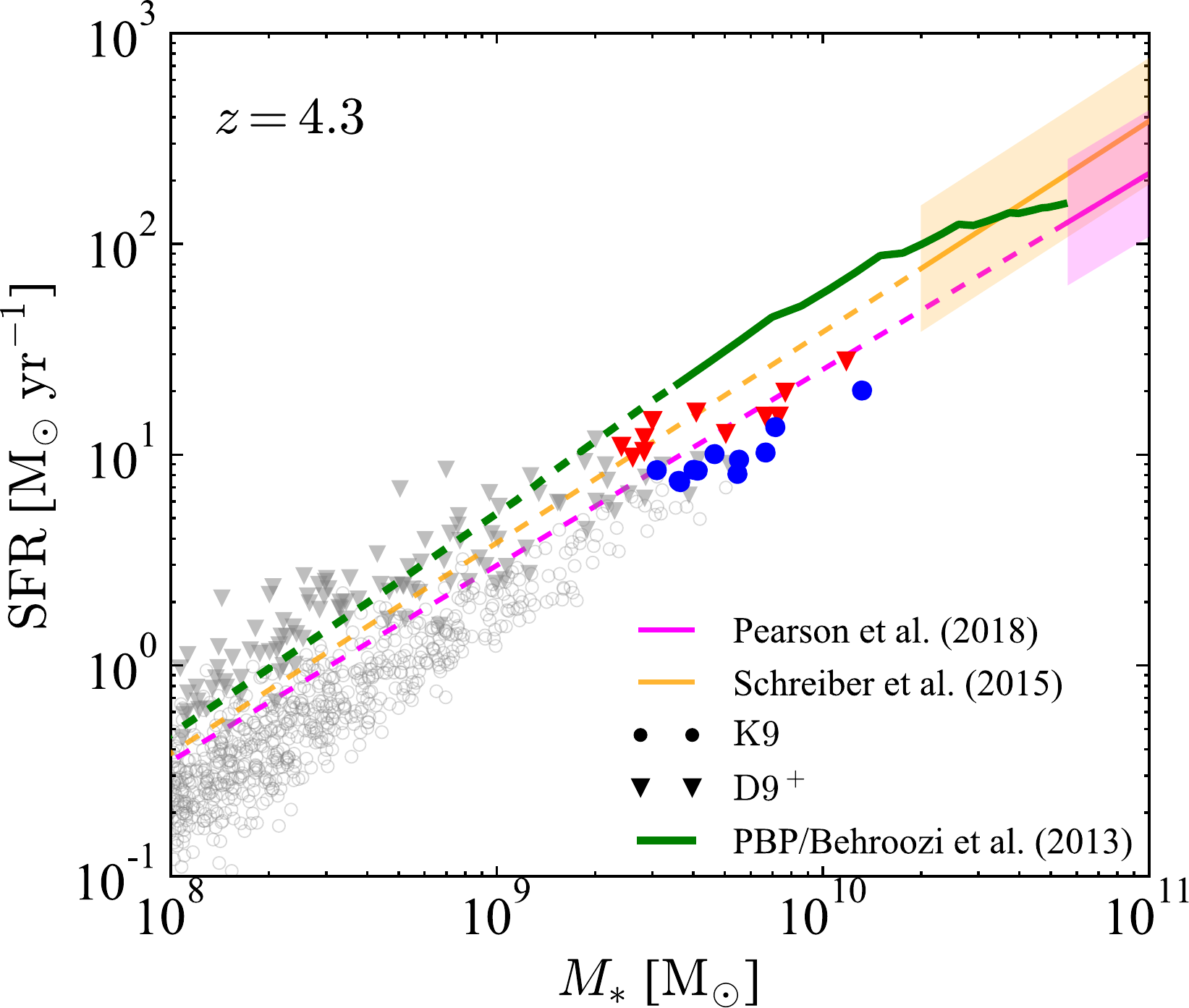}
  \caption{SFR versus stellar mass for the galaxies of the D9$^+$ and K9 runs at $z \approx 4.3$ (grey symbols, coloured ones highlight the 11 galaxies selected in  section~\ref{sec:dynresults}). 
  By construction, galaxies in the PBP model lie along the green solid curve derived in \citet{behroozi+13}.
  For comparison, we also show fits to
  the  main sequence of star-forming galaxies derived from observations at similar redshifts \citep{schreiber+15, pearson+18}. Solid lines are used
  in the range of $M_*$ probed by observations. Dashed lines, instead, indicate the extrapolation of the fits to lower stellar masses. The shaded areas denote the intrinsic scatter measured about the main sequence.}
	\label{fig:main_sequenc}
\end{figure}

In all cases, the simulated galaxies match well the properties and the evolutionary trends of the less extreme observed objects. This indicates good agreement. In fact, given the relatively small size of our simulation boxes, our synthetic galaxies can only be representative of the typical galaxy population and not of the tails of the distributions sampled by current observations.
Fig.~\ref{fig:main_sequenc} provides evidence in this direction.
Here we show a scatterplot of the SFR against stellar mass for the galaxies in the D9$^+$ and K9 simulations at $z\approx 4.3$.
These quantities are tightly correlated. The points in the plot align in a similar fashion to the observed main sequence of star-forming galaxies \citep[e.g.][]{schreiber+15, pearson+18}. Interestingly, the sequence in the simulation nicely extends to low stellar masses that are not probed by current observations. Note that the galaxies of the PBP model are assumed to lie along the green solid curve
that relates the average SFR and the average $M_*$ for different halo masses\footnote{Contrary to what happens at lower and higher redshift,
data at $z\sim 4$ constrain this relation only for $M_*\gtrsim {\mathrm {a\ few}}\times 10^9$ M$_\odot$ or, equivalently,
$M_{\mathrm h}\gtrsim {\mathrm {a\ few}}\times 10^{11}$ M$_\odot$ \citep{behroozi+13}.}.

\begin{figure*}
  \includegraphics[width=\linewidth]{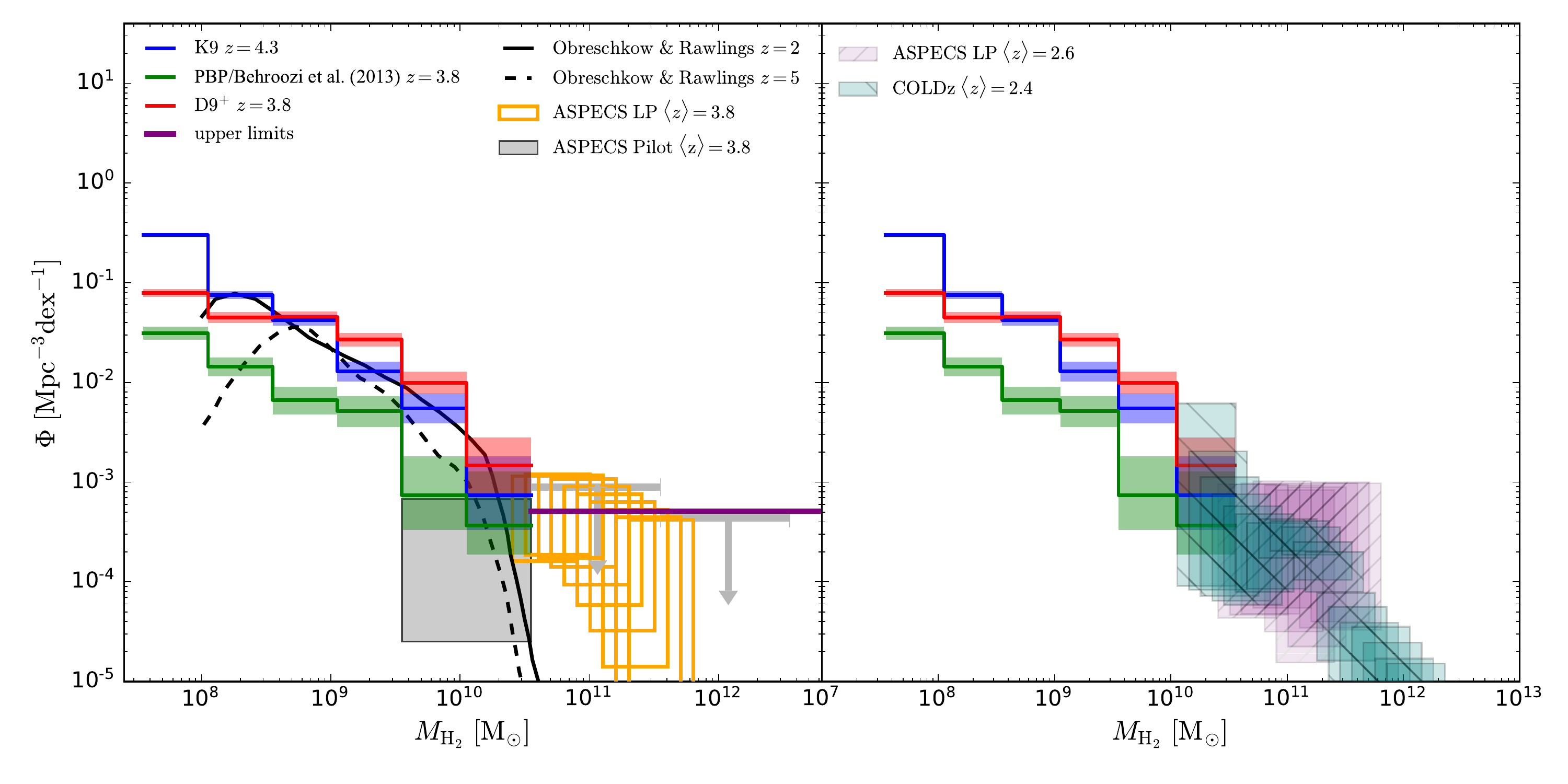}
  \caption{Left: the \hh~mass functions (and their 1-$\sigma$ Poisson uncertainties) derived at a mean redshift  of 3.8 (the full range is $3.01<z<4.48$) from the  CO(1-0) luminosity function in the ASPECS Pilot program \citep[gray box and downwards arrows,][]{decarli+16}, 
  and in the ASPECS Large Program 3mm data \citep[gold-framed sliding boxes,][]{decarli+19} are compared with the results of our simulations (coloured histograms with errorbars that include Poisson errors and sample variance). The horizontal purple line indicates the upper limit for the MF corresponding to zero counts in the simulations in bins of 0.5 dex.  The solid and dashed black curves   show the models from \citet{obreschkow&rawlings09} at $z=2$ and $z=5$. Right: lower-redshift data from the ASPECS Large Program (hatched purple boxes) and COLDz \citep[hatched teal frames,][]{riechers+18} are compared with our simulation results (same as in the left-hand panel).}
  \label{fig:cosmic_h2_MF}
\end{figure*}

\subsection{\hh ~mass function}
\label{sec:mh2}

Studying the evolution of the \hh~mass function (MF) provides a convenient way to express information about the molecular content of the Universe. This quantity gives the comoving number density of galaxies per unit $M_{\mathrm H_2}$ interval, i.e. ${\mathrm d}n=\phi\,{\mathrm d}M_{\mathrm H_2}$.
As commonly done in the high-redshift literature, we use here the MF per unit log-mass interval
\begin{equation}
\Phi(M_{\mathrm H_2},z)=\frac{{\mathrm d}n}{{\mathrm d} \log M_{\mathrm H_2}}=
\frac{M_{\mathrm H_2}}{\log e}\,\phi(M_{\mathrm H_2},z)\;.
\end{equation}
Observationally, the MF is estimated from the CO luminosity function adopting a CO-to-\hh~conversion factor inferred for normal galaxies in the local Universe or for near-infrared galaxies at intermediate redshifts. 
At high $z$, data are still scarce and only constrain the high-mass end of the MF.
In the left-hand panel of Fig.~\ref{fig:cosmic_h2_MF}, we report two recent estimates based on the ALMA Spectroscopic Survey in the Hubble Ultra Deep Field (ASPECS)\footnote{All the ASPECS results assume the CO[$J$-$(J-1)$]-to-CO(1-0) luminosity ratios derived by \citet{daddi+15} for normal star-forming galaxies, a CO(1-0)-to-\hh~conversion factor of $\alpha_{\mathrm{CO}} = 3.6$ M$_\odot$ (K km s$^{-1}$ pc$^2$)$^{-1}$
, as well as a cosmological model with $\Omega_{\mathrm m}=0.3$, $\Omega_\Lambda=0.7$ and $H_0=70$ km s$^{-1}$ Mpc$^{-1}$.} for sources in the redshift range $3.01<z<4.48$ with average $\langle z \rangle = 3.8$ \citep{decarli+16,decarli+19}.
The grey box represents the measurement of the ASPECS Pilot program at $M_{\mathrm H_2}\sim 10^{10}$ M$_\odot$ while the downward arrows denote the corresponding upper limits at larger $M_{\mathrm H_2}$ \citep{decarli+16}.
On the other hand, the gold-framed sliding boxes depict the results from the ASPECS Large Program (LP) 3mm data in the same redshift interval \citep{decarli+19}.
Superimposed, we plot the MF obtained from our simulations at $z=3.8$ for D9$^+$ (red histogram) and PBP (green) runs and at $z=4.3$ for the K9 (blue) run. In this case, we consider only central galaxies\footnote{Section~\ref{sec:results} shows that, for low halo masses, $M_{\mathrm H_2}$ is affected by the limited spatial resolution of the K and D simulations. Therefore, the corresponding MFs could be underestimated at the low-mass end.} (no satellites) with \mhh $\geq 4 \times 10^7 \Mo$. 
Error bars (the shaded regions) include the contributions from Poisson noise and sample variance (which we estimate from the two-point correlation function of the simulated galaxies).
The horizontal purple line indicates the upper limit corresponding to counting zero galaxies in each bin of the MF extracted from the simulations.
The PBP model provides the best agreement with the ASPECS Pilot data while the dynamical and equilibrium models roughly predict two to three times higher counts. On the other hand, all simulations are 
compatible with ASPECS LP, although the \mhh ~ranges probed by the data and models do not overlap\footnote{\citet{decarli+19} locate the $5\sigma$-detection limit
for ASPECS LP at approximately $2.7\times 10^{10}$ M$_\odot$
(under a series of assumptions listed in their Table 1) 
which nearly coincides with the \hh~mass of the most massive object
in our simulation box.} given the relatively small size of our computational volume.
For reference, we also show the MFs calculated at $z=2$ and 5 by \citet{obreschkow&rawlings09} assuming a relation between the interstellar gas pressure and the local molecular fraction \citep{obreschkow+09}. These predictions were obtained by post-processing  the semi-analytic galaxy catalogue of \citet{delucia&blaizot07}. 
They lie in the same ballpark as our simulations and are somewhat intermediate between the results of the PBP and D9$^+$ runs.

The mass function is often approximated by a Schechter function
\begin{equation}
    \phi=\frac{n_*}{M_*}\,\left(\frac{M_\mathrm{H_2}}{M_*}\right)^\alpha \exp\left(-\frac{M_\mathrm{H_2}}{M_*}\right)\;,
\end{equation}
or, equivalently,
\begin{equation}
    \log \Phi =(\alpha+1)\,\log \frac{M_\mathrm{H_2}}{M_*}-\frac{M_\mathrm{H_2}}{M_*}\,\log e+\log n_*-\log \log e\;.
\end{equation}
Here, $n_*$ is a normalisation constant, $\alpha$ indicates the low-mass slope, and $M_*$ denotes the knee of the mass function, above which galaxy counts fall off exponentially. 
Fitting a Schechter funtion to the ASPECS LP data gives $M_*\simeq 10^{10.5}$ M$_\odot$ \citep{decarli+19}. On the other hand, no robust constraints can be set on $\alpha$ that turns out to be very sensitive to the corrections applied for fidelity and completeness \citep{decarli+19}.
All the models displayed in Fig.~\ref{fig:cosmic_h2_MF} suggest that the ASPECS measurements should indeed sit around the knee of the MF.
In the PBP model, the cutoff is located at $M_*\simeq 10^{9.5\mathrm{-}10}$ M$_\odot$ while it is shifted up by approximately half a dex in the K9 and D9$^+$ runs.
Another interesting aspect worth mentioning is that the faint-end slope in the D9$^+$ run ($\alpha\simeq -1.3$) is substantially shallower than in the other two simulations ($\alpha \simeq -1.9$ for K9 and $\alpha \simeq -1.7$ for PBP).
Fairly flat low-mass slopes (at least down to $\simeq 10^{10}$ M$_\odot$)
have been measured at $z\simeq 2-3$ by ASPECS LP and COLDz \citep{riechers+18}
-- see the right-hand panel in Fig.~\ref{fig:cosmic_h2_MF}. The values of $\alpha$ found in the simulations imply that most of the \hh~in the cosmos sits within reservoirs with $M_{\mathrm H_2}$ slightly smaller than $M_*$ and thus just a bit below the current detection limits. 
For instance, if the mass function closely follows a Schechter function with $\alpha=-1.3$ ($-1.9$), then only 28 (16) per cent of the total \hh~lies within objects with $M_{\mathrm H_2}>M_*$ while already 84 (66) per cent is found in galaxies with $M_{\mathrm H_2}>0.1 M_*$. We will further discuss this in the next section.

\subsection{Cosmic \hh~density}
\label{sec:cosmic-h2}

\begin{figure*}
  \includegraphics[width=\linewidth]{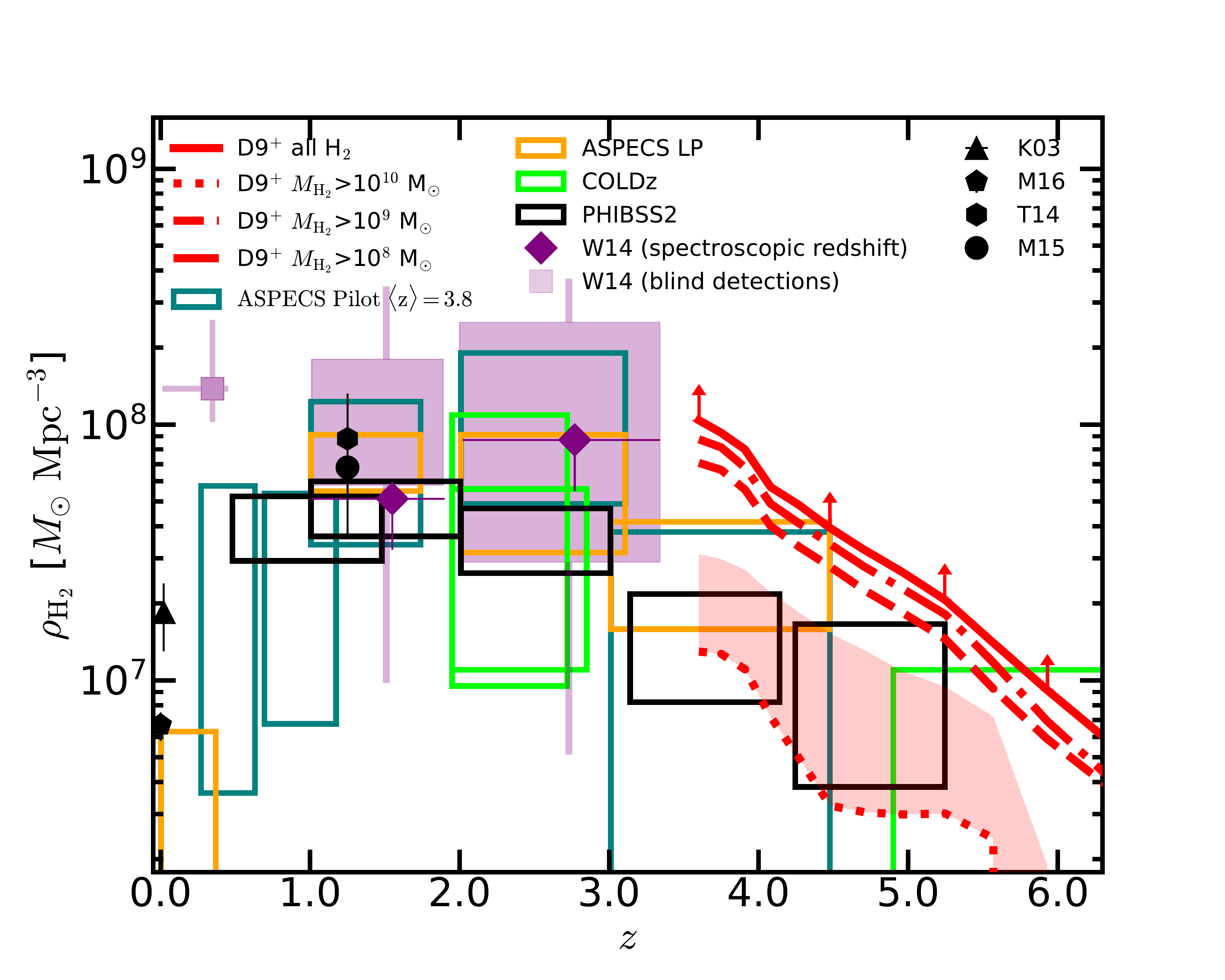}
  \caption{The redshift evolution of the cosmic $\rm{H}_2$ mass density in the \dynp~simulation (thick red curves with different line styles corresponding to various minimum galactic \hh~masses as indicated by the labels) is compared with the observational constraints from \citet[][W14]{WALTER},  \citet[][ASPECS Pilot]{decarli+16}, \citet[][ASPECS LP]{decarli+19}, \citet[][PHIBSS2]{lenkic+19}, \citet[][COLDz]{riechers+18}, \citet[][K03]{keres+03}, \citet[][M17]{maeda+17}, \citet[][T14]{TOMCZAK} and \citet[][M15]{MORTLOCK}. The red-shaded area indicates the expected contribution from massive haloes that are underrepresented in our simulation box. }
	\label{fig:cosmic_h2_content_part_1}
\end{figure*}

\begin{figure*}
  \includegraphics[width=\linewidth]{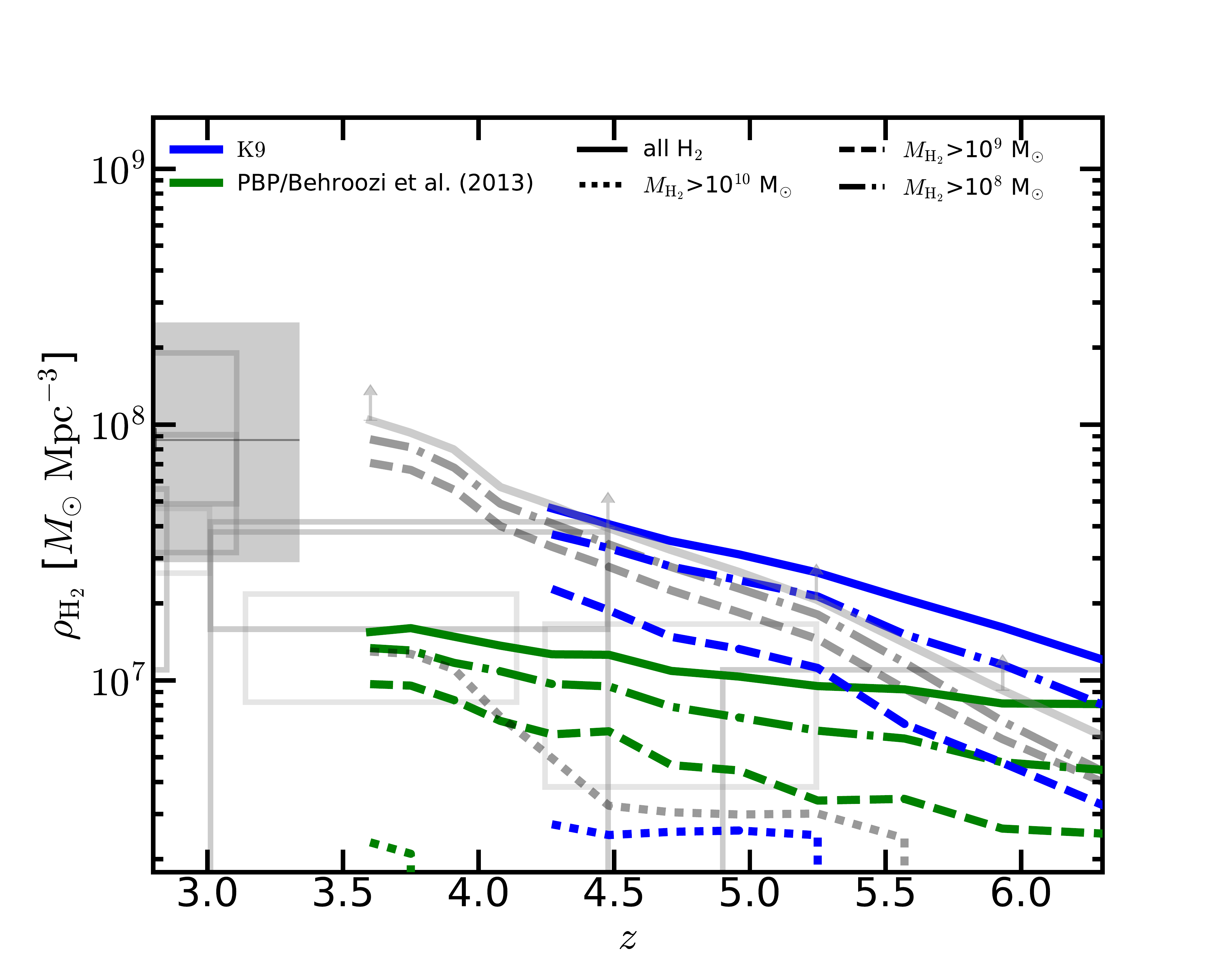}
  \caption{As in Fig.~\ref{fig:cosmic_h2_content_part_1} but for the K9 (blue) and PBP (green) simulations and focusing on redshifts $z>3$. To ease comparison,  Fig.~\ref{fig:cosmic_h2_content_part_1} is reproduced in the background with gray tones.
   }
	\label{fig:cosmic_h2_content_part_2}
\end{figure*}

Determining the redshift evolution of the cosmic \hh~mass density,
\begin{equation}
\rho_{\mathrm H_2}(z)=\int \!\!\!M_{\mathrm H_2}\,\phi\, \mathrm{d}M_{\mathrm H_2}
= \int M_{\mathrm H_2}\,\Phi \, \mathrm{d}\log M_{\mathrm H_2} \;,
\end{equation}
has been the subject of continued observational effort.
The current state of the art is summarized in Fig.~\ref{fig:cosmic_h2_content_part_1}.
Although quite noisy, the data show an evolutionary trend for which $\rho_{\mathrm H_2}$ peaks at $z\approx 2$, in good agreement with the cosmic SF history \citep[e.g.][]{madau&dickinson14}. 
Observational constraints are looser at $z>3$ as i) they are based on very small samples that only include the most massive galaxies; ii) the CO-to-\hh~conversion factors are uncertain, and iii) the intrinsic shape of the CO luminosity function is unknown. 
Given this, and the large Poisson errors, authors generally do not attempt to correct their estimates for the contribution of faint undetected objects.

Overplotted is the cosmic \hh~mass density extracted from our D9$^+$ simulation  (thick red lines). In particular, the dotted curve represents the contribution from the galaxies with $M_{\mathrm H_2}>10^{10}$ M$_\odot$. This threshold approximately matches the detection limit of current CO surveys at high redshift \citep[e.g. it lies slightly below the
$5\sigma$-detection limit for ASPECS LP,][]{decarli+19}.
For a proper comparison of the numerical results with the observations, we need to account for galaxies hosted by the rare, very massive haloes that are unlikely to form in our relatively small simulation box. To estimate their overall contribution to the cosmic \hh~density, we proceed as follows. For the most massive haloes at $4\lesssim z \lesssim 6$, we find that \mhh$\simeq M_{\mathrm h}/30$  (see e.g. Fig.~\ref{fig:dyn_models}). Therefore, we first compute the total mass density contributed by haloes that are more massive than those appearing in the simulation using a fit for the halo mass function \citep{2001MNRAS.323....1S, 2008ApJ...688..709T}.
We then rescale the result by a factor of 30 to get a rough estimate of the corresponding \hh~density. The final contribution of the `missing' haloes is shown as a shaded region lying above the dotted thick red line in Fig.~\ref{fig:cosmic_h2_content_part_1}.
At $3.5 \lesssim z\lesssim 5.5$, the shaded area and the dotted line give nearly equal contributions. 
By considering their sum, we conclude that
the agreement between the simulation and the observations is very good.

We now focus on the low-mass end.
The dashed and dash-dotted curves in Fig.~\ref{fig:cosmic_h2_content_part_1} represent the contribution from galaxies with $M_{\mathrm H_2}>10^{9}$ M$_\odot$ and $M_{\mathrm H_2}>10^{8}$ M$_\odot$, respectively.
Finally, the solid curve accounts for the total \hh~mass in the computational volume (without correcting for the missing haloes). Since the simulation likely underestimates the molecular mass of the galaxies residing in low-mass haloes (see section \ref{sec:dynresults}), we have represented this result with upward-pointing arrows to indicate that it is likely a lower limit.
Our results suggest that current measurements of the \hh~mass density may be underestimated by a redshift-dependent factor that ranges between 2 and 3 at $3.5\lesssim z \lesssim 5.5$ (after taking into account the contribution from the massive haloes that are underepresented in our box).
The galaxies that host the undetected molecules
should also contribute an important fraction of the cosmic SFR.

The redshift evolution of the cosmic \hh~density in the PBP and K9 simulations is presented in Fig.~\ref{fig:cosmic_h2_content_part_2}.
The K9 predictions are largely consistent with those of the D9$^+$ run.
However, they show less evolution at low $M_{\mathrm H_2}$ as low-mass molecular reservoirs are in place earlier in this model (see the top-left panel in Fig.~\ref{fig:results_panel}).
On the other hand, the PBP model predicts a milder redshift evolution of the cosmic \hh~density between redshift 6.5 and 3.6 with respect to the other runs. The total \hh~density at $z=3.6$ is nearly an order of magnitude lower than in the D9$^+$ run. When one takes into account the
current detection limits, it appears difficult to reconcile the PBP results with the measurements from the IRAM Plateau de Bure HIgh-z Blue Sequence Survey 2 \citep[][]{lenkic+19} and ASPECS LP \citep{decarli+19}.

Another interesting aspect of the cosmic \hh~density, noticeable in Fig.~\ref{fig:cosmic_h2_content_part_2}, is that the spacing between
the lines drawn with different styles varies among the different runs but, barring the PBP model, does not change much with redshift.
This behaviour reflects the different shapes of the \hh~MF in the three simulations. Therefore, Fig.~\ref{fig:cosmic_h2_content_part_2}
extends the analysis presented in Fig.~\ref{fig:cosmic_h2_MF}
to a broader redshift range.

\section{Summary}
\label{sec:summary}

In this paper, we have analyzed and compared three approximate methods for tracking the formation and evolution of \hh~in cosmological simulations of galaxy formation.
The first, dubbed PBP, is a semi-empirical model that associates a \hh~mass to a galaxy based on the mass and the redshift of its host dark-matter halo \citep{popping+15}.
The second, labelled KMT, assumes chemical equilibrium between the \hh~formation and destruction rates \citep{krumholz+09}. 
The third, called DYN, fully solves the out-of-equilibrium rate equations and accounts for the unresolved structure of molecular clouds using a sub grid model \citep{tomassetti+15}.
Both the KMT and the DYN models require as an input local estimates of the density, metallicity and the intensity of radiation in the LW band. We compute this last quantity by propagating radiation from the stellar particles in the simulations.
Furthermore, in the simulations based on the KMT model, we link SF to the local density of cold gas, whereas we use the \hh~density in conjunction with the DYN model. 

Each of the algorithms listed above represents a broad class of models (semi-empirical, equilibrium, non-equilibrium) that have been adopted in the literature, sometimes with different implementations.
For the scientific applications, however, different authors have used wildly different spatial resolutions making it difficult to compare their results.
Therefore, as a first task, we have investigated how the finite spatial resolution of the numerical simulations impacts the predictions of the \hh~content of the synthetic galaxies in the KMT and DYN models. 
Using the \textsc{Ramses} code, we have run a suite of simulations (four for DYN model and three for the KMT one) that reach different maximum levels of refinement for the dark-matter and gaseous components.
All the simulations start from ICs that have the same amplitudes and phases for the mutually resolved Fourier modes so that we can easily cross match the host haloes of the galaxies in the different runs.
Finally, we have compared our synthetic galaxies to a compilation of recent observational results.
Our results can be summarized as follows. 

i) The conversion rate of HI into \hh~depends on the local metallicity and density of the ISM. It turns out that the \hh~formation time can be far longer than the age of the Universe in low-mass high-$z$ galaxies with low $Z$ \citep{kuhlen+12,jaacks+13, kuhlen+13, thompson+14, tomassetti+15}.
However, in Fig.~\ref{fig:cells_star_n_circle}, we have shown that the precise timing of the end of this process is resolution dependent.
At higher spatial resolutions, the progenitors of a galaxy
can reach higher gas densities and start forming molecules, stars and metals earlier than in lower-resolution runs.
Therefore, \hh~masses that are stable with respect to resolution changes can only be obtained: a) for the objects that have reached a sufficient level of metal enrichment and where the \hh-formation timescale is  shorter than the age of the Universe (see Fig. \ref{fig:cells_star_n_circle}); and b) for the galaxies that are still not forming any \hh. 
In our simulations with resolution elements of 50-100 pc at $z=3.6$, the first group corresponds to $M_{\mathrm H_2}\gtrsim 6\times 10^{9}$ M$_\odot$ and the second to
$M_{\mathrm H_2}\ll 10^{8}$ M$_\odot$ (see Fig. \ref{fig:selec_11}).

ii) On average, in our highest-resolution runs, the KMT and the DYN models generate very similar $M_{\mathrm H_2}-M_{\mathrm h}$ relations, while the PBP one produces significantly less \hh~at fixed halo mass (by a factor of 5-6, see Fig.~\ref{fig:all_models}). On an individual basis, the KMT- and DYN-based runs yield similar \hh~masses for metal-enriched galaxies with $Z\gtrsim {\mathrm{a\ few}}\times 10^{-2} \, \mathrm{Z_\odot}$ \citep[see also][]{krumholz&gnedin11}, although the values of $M_{\mathrm H_2}$ tend to be slightly higher in the DYN model, on average (see Fig. \ref{fig:results_panel}). 
Note, however, that a given galaxy gets metal enriched a later time in the DYN model.

iii) The detailed morphology of the synthetic galaxies is influenced by the maximum spatial resolution of the simulations and by the SF law (gas vs. \hh~based) that determines where and when stellar feedback injects energy into the ISM (see Fig.~\ref{fig:h2_maps}). 

(iv) The simulated galaxies with the highest $M_{\mathrm H_2}$ (whose molecular content does not depend much on spatial resolution effects) match well the properties (star formation rates, stellar and \hh~masses) and the evolutionary trends of the less extreme objects extracted from a compilation of recent observational data (Fig.~\ref{fig:results_panel}). This is  satisfactory in all respects as current observations at high-$z$ tend to pick extreme objects that are not representative of the overall population while our relatively small simulation box targets average galaxies that populate the main sequence of star-forming objects (Fig.~\ref{fig:main_sequenc}).
Interestingly, all the synthetic galaxies that host large molecular reservoirs have disc morphologies and present spiral arms in the higher-resolution runs.
Overall,  the  differences  between  the three \hh~models are smaller than the scatter among the individual galaxies (Fig.~\ref{fig:results_panel}). Some systematic trends are noticeable, however. First, the dynamical model predicts a delayed assembly of the molecular and stellar masses. Second, the PBP model yields lower molecular and stellar masses at $z<6$. 

(v) The \hh~MFs extracted from the simulations at $z\approx 4$ show similar normalisations and cutoff scales ($M_{\mathrm H_2}\approx 10^{10}$ M$_\odot$) as recent observational data and semi-analytic models (Fig.~\ref{fig:cosmic_h2_MF}). 
The PBP model generates lower counts at fixed $M_{\mathrm H_2}$ (by a factor of $\sim 3-10$) with respect to the KMT and DYN ones. In all cases, the low-mass slopes of the MF
are sufficiently flat that only a small fraction of molecular material is contained in galaxies with $M_{\mathrm H_2}\ll 10^{10}$ M$_\odot$.

(vi) When we account for the detection limits of current CO surveys, the cosmic \hh~mass density extracted from our simulations based on the KMT or DYN models matches well with recent observations at $z>3$ (Fig. \ref{fig:cosmic_h2_content_part_1} and  \ref{fig:cosmic_h2_content_part_2}). However, our results suggest that most molecular material at high-$z$ lies yet undetected in reservoirs with $10^9<M_{\mathrm H_2}<10^{10}$ M$_\odot$. Adding the integrated contribution of these sources to the current estimates for $\rho_{\mathrm H_2}$ increases its value by a redshift-dependent factor ranging between 2 and 3. 
Similarly, they should be responsible for an important fraction of the cosmic SFR at $3.5<z<6$.
Based on our simulations,the dominant contribution to the missing \hh~comes from galaxies that lie just below the detection limits of current CO surveys and should become available
with longer integration times and/or to future facilities like the next generation Very Large Array \footnote{\url{https://ngvla.nrao.edu/}.}.


\section*{Acknowledgements}
We thank the anonymous referee for comments and suggestions that helped improving the presentation of our results. We thank Romain Teyssier for making the \textsc{Ramses} code available and for helpful discussions. We are also grateful to Laura Lenkic for kindly providing us with the observational data from PHIBSS2.
This work was supported by the Deutsche Forschungsgemeinschaft (DFG) through the Collaborative Research Center 956 {\it Conditions and Impact of Star Formation}, sub-project C4. ADL acknowledges financial supported from the Australian Research Council (project Nr. FT160100250). The authors acknowledge that the results of this research have been achieved using the PRACE-3IP project (FP7 RI-312763), using the computing resources (Cartesius) at SURF/SARA, The Netherlands. \\

\section*{Data availability}
The data underlying this article will be shared on reasonable request to the corresponding author.


\bibliographystyle{mnras}
\bibliography{references} 


\bsp	
\label{lastpage}
\end{document}